\documentclass[reprint,amsmath,amssymb,aps,prc]{revtex4-1}
\usepackage{hyperref}
\usepackage{graphicx}   
\usepackage{dcolumn}   
\usepackage{bm}        

\usepackage{hyperref}

\begin{document}
\title{Physics Program for the STAR/CBM eTOF Upgrade}

\author{The STAR Collaboration\\
 The CBM Collaboration eTOF Group}

\date{\today}

\begin{abstract}
\begin{center}
\includegraphics[width=6.0 in]{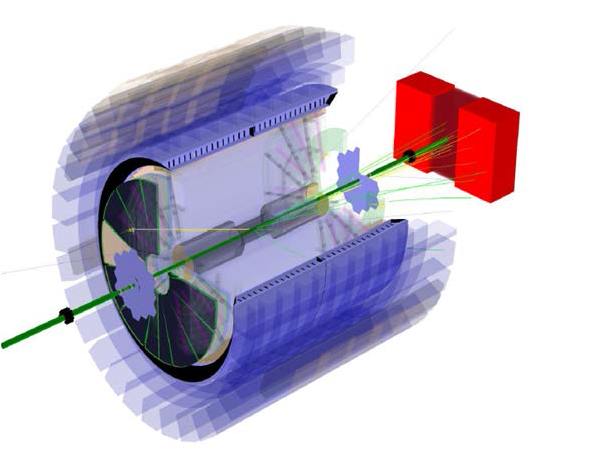}
\includegraphics[width=6.0 in]{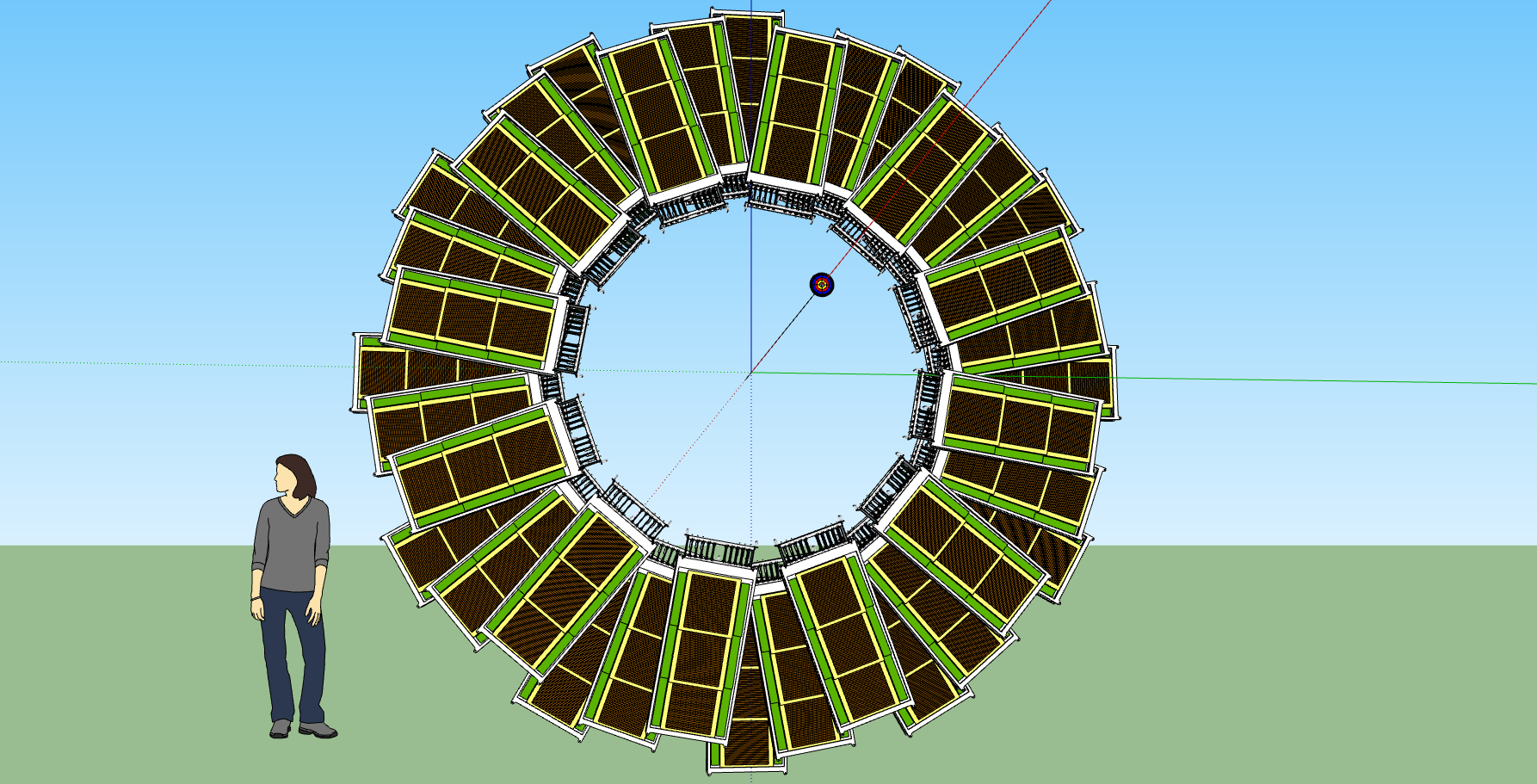}
\newpage
\textbf{ABSTRACT}
\end{center}

The STAR Collaboration and the CBM Collaboration institutions: Heidelberg, Darmstadt, CCNU,
Tsinghua, and USTC propose to install an end-cap time-of-flight upgrade (eTOF) to the STAR detector 
for the RHIC beam energy scan phase II (BES-II) program in 2019 and 2020. BES-II will cover 
the collision energy range 3.0 to 19.6 GeV. This is the region of interest in the search for a 
critical point and first-order phase transition, identified by the results from BES-I and by 
model calculations. For the collider-mode portion of the energy scan, 7.7 to 19.6 GeV, eTOF will extend particle identification (PID) for pions, kaons, and protons to a rapidity of 1.2, complementing the inner Time Projection Chamber (iTPC) upgrade to the forward tracking. The rapidity coverage for PID would extend to only 0.8 without the eTOF upgrade. The eTOF upgrade will enable precision studies of the key bulk property observables, essential to the BES-II search.
An internal fixed-target program will allow the energy scan to cover 3.0 to 7.7 GeV. The eTOF upgrade will provide essential mid-rapidity PID for the 4.5 to 7.7 GeV portion of the scan in fixed-target mode.
Otherwise there would be a large energy gap in the middle of the BES-II
program. A full description of the physics provided by the eTOF upgrade to STAR is presented in this note.

\tableofcontents

\end{abstract}

\maketitle

\section{Introduction}
The first RHIC Beam Energy Scan (BES-I) was an initial survey in which data were acquired 
from Au+Au collisions at $\sqrt{s_{NN}}$ = 62.4, 39, 27, 19.6, 14.5, 11.5, and 7.7 GeV 
in 2010, 2011, and 2014~\cite{BES}. The results from that program have been used to develop
a focused BES-II program, scheduled to run in 
2019 and 2020~\cite{BESII}. The BES-II program will rely on low-energy electron cooling 
and longitudinally extended bunches to improve the luminosity~\cite{Cool09}. 
The program in normal collider mode will cover
the energy range from 7.7 to 19.6 GeV where the most promising results from 
BES-I were seen. The energies from 3.0 to 7.7 GeV will be accessible through the use of an internal 
fixed target~\cite{FXT}. 
Major upgrades to the STAR detector between BES-I and BES-II will allow for more refined studies. 
This document presents the benefits of the addition of an end-cap time-of-flight 
system (eTOF). The eTOF upgrade will provide particle identification (PID) in the extended
pseudorapidity range provided by the iTPC upgrade~\cite{iTPC} to the main tracking 
chamber~\cite{TPC}.

\begin{figure}
 \begin{center}
 \includegraphics[width=0.45\textwidth]{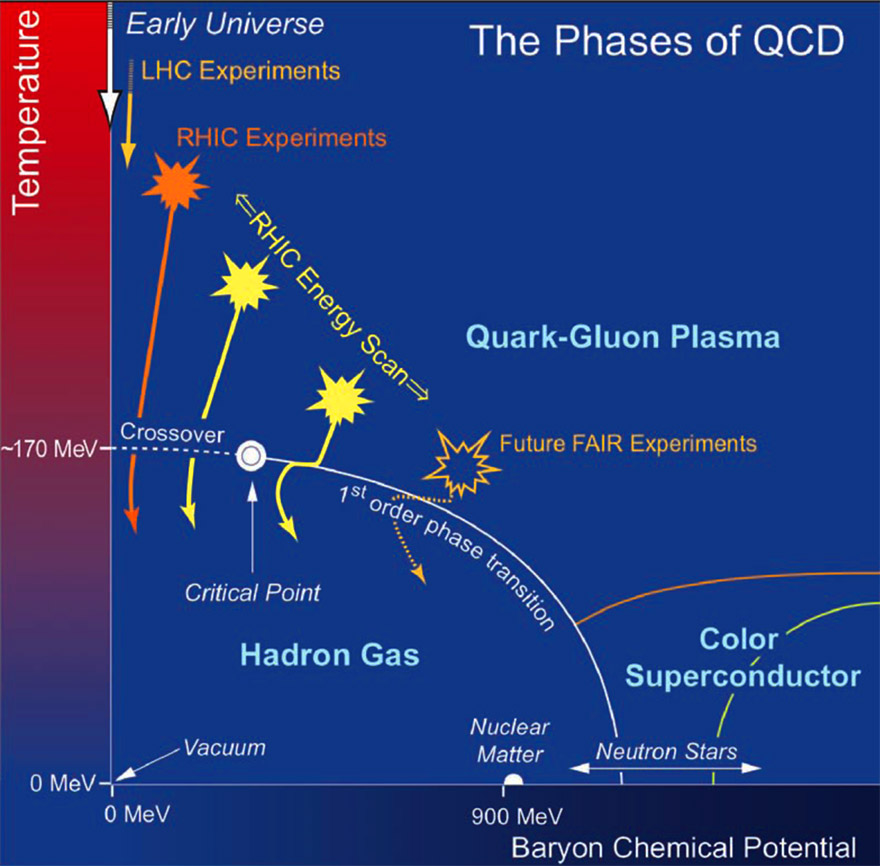}
 \caption{A conjectured QCD phase diagram with boundaries that define various 
states of QCD matter~\cite{Ullrich}.} 
 \label{Phase_Diagram}
 \end{center}
\end{figure}

The BES-II program is designed to study the phase diagram of QCD matter 
(see Fig.~\ref{Phase_Diagram}). The
program has several goals:
\begin{itemize} 
\item{To determine the temperature ($T$) and baryon chemical potential ($\mu_B$) at chemical freezeout 
for $Au+Au$ events at the beam energy where the onset of deconfinement occurs. This
would establish the basic structure of the QCD phase diagram.}
\item{To seek evidence of the softening of the equation of state, consistent with a first-order 
phase transition, to understand the nature of the phase boundary.}
\item{To look for enhanced fluctuations, which are a signature of critical behavior, to localize the 
possible critical point should the phase boundary change from a first-order to a crossover transition.}
\item{To observe the in-medium modification of the light vector meson mass, 
quantifying the effect of chiral symmetry restoration at high baryon
densities.
}
\end{itemize}

For the collider part of the program, the upgrades extend 
the pseudorapidity coverage with PID from $|\eta|<1.0$ to $|\eta|<1.5$. The eTOF is needed for 
PID at forward rapidities because the $p_Z$ boost moves many particles beyond the limits of 
PID based on energy loss, $dE/dx$. 
This extended coverage will allow for rapidity-dependent studies of the
key physics observables, important because the partial stopping of the incident nucleons 
changes the nature of the system, effectively creating a gradient in $\mu_B$ as a function of rapidity. 
The iTPC and eTOF upgrades will benefit the
fixed-target program in a different way. 
In fixed-target collisions, the center-of-mass is boosted in rapidity as a function of beam energy. 
Mid-rapidity will still fall inside the main TPC/TOF acceptance window for $\sqrt{s_{NN}}$ = 
3.0 to 4.5 GeV. However, the additional coverage of the iTPC/eTOF is needed for $\sqrt{s_{NN}}$ = 
4.5 to 7.7 GeV to provide mid-rapidity PID. The iTPC/eTOF upgrades are essential to allow a complete,
gap-free scan from 3.0 to 19.6 GeV in the combined fixed-target and collider
program of BES-II.

\section{eTOF Improvements to the Physics of the BES-II Collider Program}
\subsection{Acceptance}
\begin{figure}[th]
\includegraphics[scale=0.4]{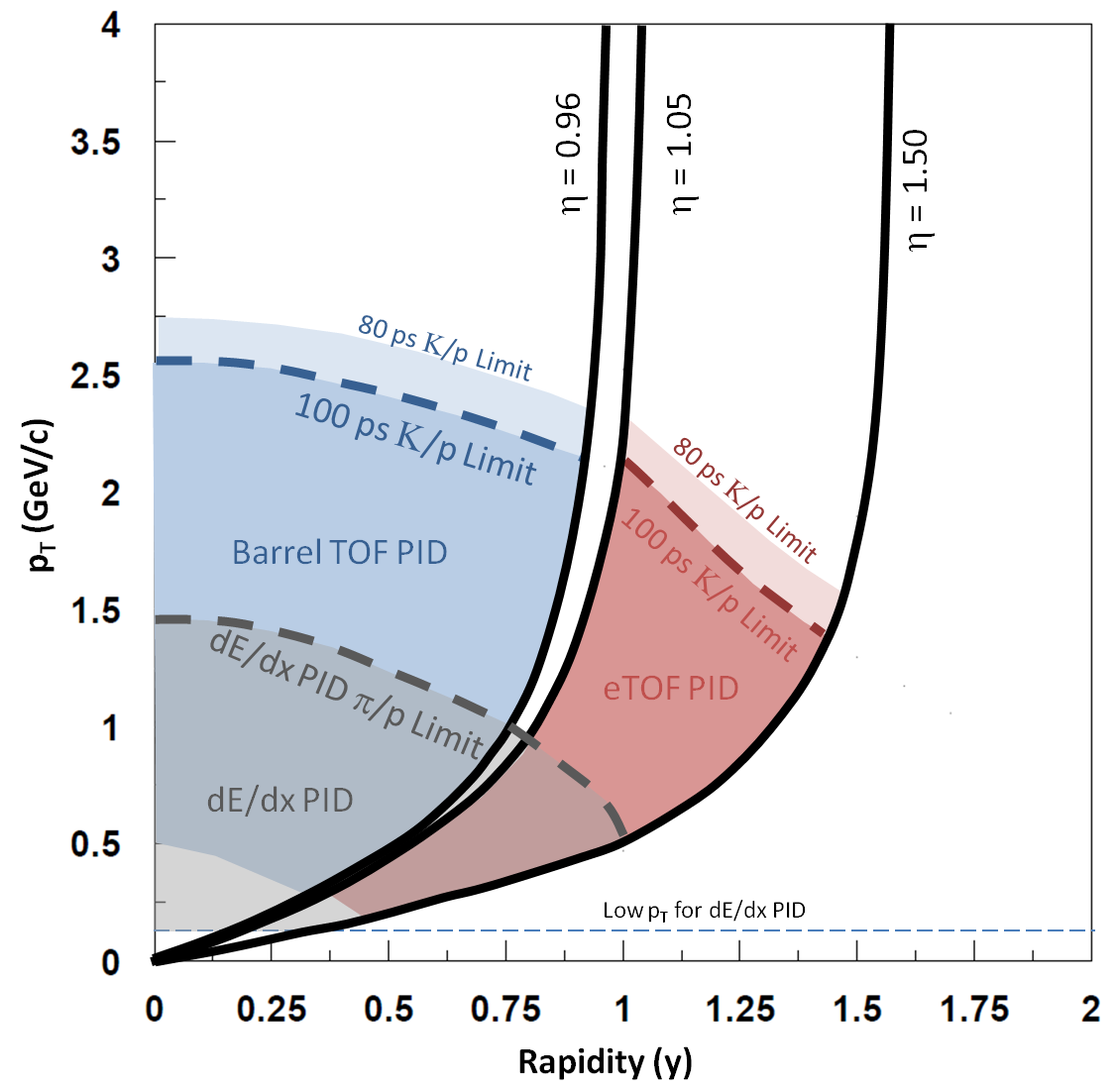}
\caption{The $p_T-y$ acceptance map for protons showing the limits due to
tracking coverage and PID.}
\label{Acceptance_pro}
\end{figure}

\begin{figure}[th]
\includegraphics[scale=0.4]{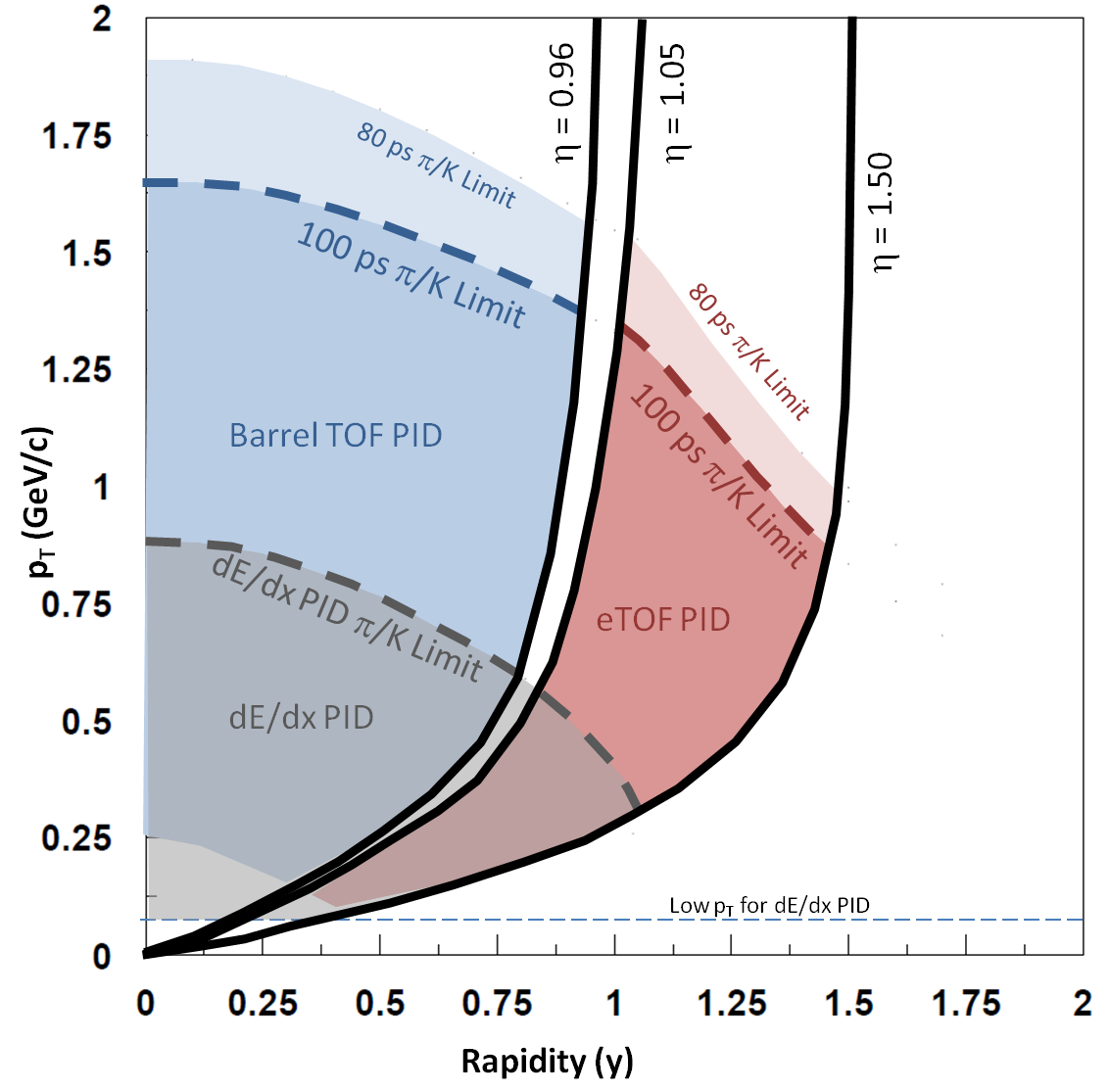}
\caption{The $p_T-y$ acceptance map for kaons showing the limits due to
tracking coverage and PID.}
\label{Acceptance_kap}
\end{figure}

\begin{figure}[th]
\includegraphics[scale=0.4]{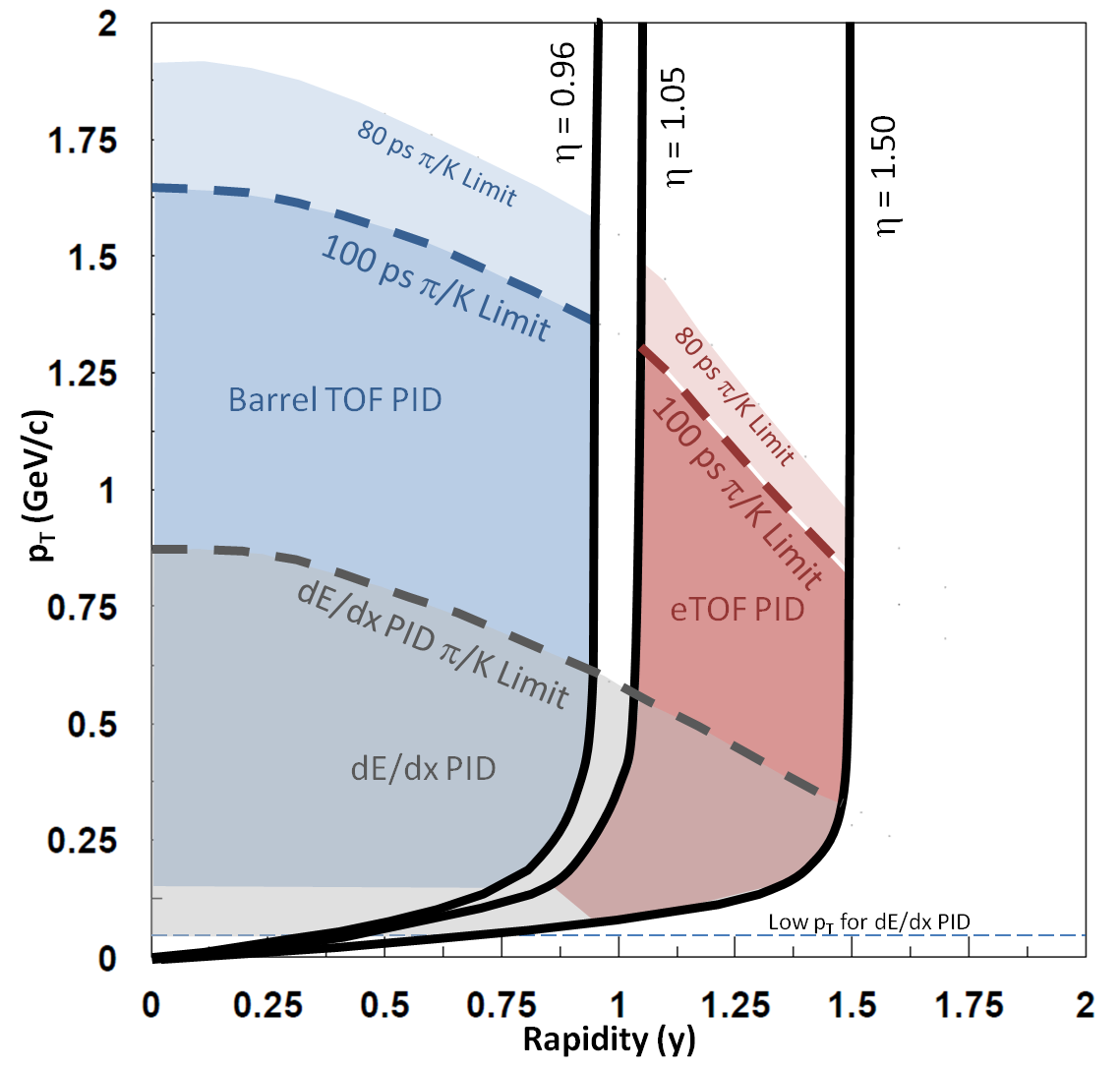}
\caption{The $p_T-y$ acceptance map for pions showing the limits due to
tracking coverage and PID.}
\label{Acceptance_pip}
\end{figure}

The nature of the improvements to the physics reach of the BES-II program is
dependent on the details of the improved acceptance. There are four key areas 
which are extended by the iTPC and eTOF detector upgrades: 
\begin{itemize}
\item{the low-$p_T$ acceptance,}
\item{the pseudorapidity coverage,} 
\item{the $dE/dx$ PID limits,} 
\item{the TOF PID limits.} 
\end{itemize}
The transformation
Jacobian from pseudorapidity to rapidity is different for each particle species.  
The $\eta$ coverage limits, which are determined by the hardware, have to be converted 
to $y$ using the appropriate transformation Jacobians. 
Because different species overlap in different PID spaces, a separate $p_T$ vs. $y$ 
acceptance map showing the $\eta$ tracking coverage limits 
must be generated for each particle species: $\pi$, $K$, and $p$ 
(see Figs.~\ref{Acceptance_pro}, \ref{Acceptance_kap}, and \ref{Acceptance_pip}).

The low-$p_T$ acceptance limit is defined by the tracking detector, while the low-$p_T$ 
TOF PID limit is set by the material of the detector and the magnetic field. For acceptance, 
a track must be identified
and associated with the primary vertex. The pattern recognition in the STAR track
finding software requires a minimum of ten hits, which sets the absolute low-$p_T$ limit 
of 50 MeV/c. However, the effect of multiple scattering
and projection back to within 3 cm of the primary vertex sets effective limits of 125 MeV/c
for protons and 65 MeV/c for kaons at mid-rapidity. The pion multiple scattering limit is 
20~MeV/c, and therefore multiple scattering does not set the acceptance limit for pions. 
These multiple scattering limits are rapidity-dependent as the material through which a
track must pass and the projection length from the inner pad rows to the primary vertex 
increase as $1/\cos\theta$, where $\theta$ is the polar angle with respect to the beam axis. 
However, the multiple scattering is also reduced by $1/\beta$ as
one goes away from mid-rapidity. 
In the barrel TOF at $y$ = 0, low-$p_T$ TOF PID limits of 500 MeV/c for protons,
250 MeV/c for kaons, and 165 MeV/c
for pions are observed. The low-$p_T$ limit for the pions is set by the 0.5 T field of the STAR solenoid.
The limits for kaons and protons are set by the energy loss in the outer field
cage. These limitations for kaons and protons will vary with $\eta$ as the material budget increases as $1/\cos{\theta}$ and $dE/dx$ decreases as $1/\beta$. For the eTOF, the effect of the material will be more complicated. The end-cap readout region of the TPC is far less uniform than the outer field cage.  The low-$p_T$ thresholds are shown in 
Figs.~\ref{Acceptance_pro}, \ref{Acceptance_kap}, and \ref{Acceptance_pip}.

The pseudorapidity acceptance limits of the detector are defined by the tracking performance and 
by the geometric extent of the TOF arrays. The barrel TOF system provides coverage to $|\eta|$ = 0.96 which is not limited by tracking performance. The eTOF system will be mounted at a distance
of 280 cm from the center of the detector and will have a radial extent from 105 to 225 cm.
This provides geometric coverage of $1.05 < \eta < 1.7$, and
leaves a small $\eta$ gap between the two TOF systems. The limitation of at least ten hits to
find a track sets an absolute high-$\eta$ tracking limit for
the iTPC of $\eta$~=~1.7. However, for these very forward tracks, the shorter track length and
longer projection length from the inner field cage to the primary vertex reduce the tracking efficiency. The efficiency has been demonstrated to be 50\% at $\eta$~=~1.5. Optimization of the
software matching tracks to the primary vertex will allow some improvement.  
The $\eta$ tracking coverage limits are
shown as functions of $y$ and $p_T$ in Figs.~\ref{Acceptance_pro}, \ref{Acceptance_kap}, 
and \ref{Acceptance_pip}.

The STAR detector has the benefit of multiple and overlapping means of PID: $dE/dx$
and time-of-flight. 
The $dE/dx$ resolution 
of gas tracking chambers was empirically studied by Allison and
Cobb~\cite{Allison1980}. Their formula for the percent resolution is:
\begin{equation}
\sigma_{dE/dx} = 0.47 N_{\rm hits}^{-0.46}(Ph)^{-0.32}
\end{equation}
where $N_{\rm hits}$ is the number of samples, $P$ is the pressure in atmospheres, and $h$ is the 
pad height or length in cm. The outer sectors cover radii from 126-190 cm with 32 pad rows
of 1.95 cm pads. The iTPC inner sectors cover radii from 60-120 cm with
40 pad rows of 1.55 cm pads. From these pad dimensions, one can determine the tracking
length for $dE/dx$ resolution as a function of pseudorapidity. The $dE/dx$ response
as a function of momentum for each particle species is given by the
parametrizations in Ref.~\cite{Bichsel}.
Using the resolutions and the parametrized response, 
we can determine the momentum limits where pions can no longer be resolved from kaons, 
and where protons can no longer be resolved from pions.
A sample of the relevant values for PID using a 
$2 \sigma_{dE/dx}$ selection are shown in Table~\ref{dEdx_table}. These $2 \sigma_{dE/dx}$ PID limits are
shown as functions of $y$ and $p_T$ in Figs.~\ref{Acceptance_pro}, \ref{Acceptance_kap}, 
and \ref{Acceptance_pip}.

\begin{table}[h]
\caption{The track length, hits, $dE/dx$ resolution, and $2\sigma_{dE/dx}$ $p_T$ limits for PID using 
$dE/dx$ for various values of $\eta$ for the new configuration (iTPC).}
\label{dEdx_table}
\begin{tabular*}{0.5\textwidth}{@{}l*{15}{@{\extracolsep{0pt
          plus12pt}}l}}
$\eta$ & Track Length & $N_{\rm hits}$ & $\sigma_{dE/dx}$ & $\pi/K$ & $\pi/p$ \\
 & (CM) & & (\%) & (GeV/c) & (GeV/c) \\
\hline
0.0 & 126  & 72 & 5.5  & 0.88 & 1.47 \\
\hline
0.5 & 142  & 72 & 5.2  & 0.79 & 1.31 \\
\hline
1.0 & 163  & 49 & 4.8  & 0.59 & 0.97 \\
\hline
1.2 & 123  & 40 & 5.5  & 0.49 & 0.81 \\
\hline
1.5 & 80  & 24 & 6.7  & 0.35 & 0.60 \\
\hline
\end{tabular*}
\end{table}

The PID derived from TOF measurements is a function of the timing resolution ($\delta t/t$) of the TOF modules, momentum resolution ($\delta p/p$), and the resolution in the flight path ($\delta s/s$) of the particles. The mass resolution ($\delta M/M$) is given by:
\begin{equation}
\left(\frac{\delta M}{M}\right)^2 = 4\left[\left(\frac{\delta p-1}{p-1}\right)^2 + \gamma^4 \left\{\left(\frac{\delta s}{s}\right)^2 +\left(\frac{\delta t}{t}\right)^2\right\}\right]
\end{equation}
The momentum resolution of a tracking chamber is given by:
\begin{equation}
\left(\delta k\right)^2 = \left(\delta k_{\rm MS}\right)^2 + \left(\delta k_{\rm Tracking}\right)^2
\end{equation}
where $k=1/R$ ($R$ being the radius of curvature) and $p_T = 0.3qBR$ ($q$ is the charge 
and $B$ is the magnetic field). 
The multiple scattering term ($\delta k_{MS}$) is given by:
\begin{equation}
\delta k_{\rm MS} = \frac{q(0.016)}{\sqrt{sX_0}\beta}
\end{equation}
where $X_0$ is the radiation length of the material, 
and the tracking term ($\delta k_{\rm Tracking}$) is given by:
\begin{equation}
\delta k_{\rm Tracking} = \frac{\sigma_t \sqrt{720}}{s^2\sqrt{N_{\rm hits}}+4}
\end{equation}
where $\sigma_t$ is the transverse resolution of a TPC hit.
The transverse momentum resolution of the STAR TPC has been demonstrated to rise 0.85\% per GeV/c. The best resolution for protons is found to be 1.0\% at 1 GeV/c while the best resolution for 
pions is found to be 1.4\% at 0.4 GeV/c. Using these measured transverse momentum resolutions, 
the constants in the two momentum resolution terms have been fixed and used to estimate the momentum 
resolution for pions and protons at the TOF PID limits. These momentum resolutions are given in 
Table~\ref{TOF_table}.
The resolution in the flight path ($\delta s/s$) is given by the resolution of the primary 
vertex and the spatial accuracy of the TOF modules. The accuracy of the primary vertex is determined to a few hundred microns and has a dependence on multiplicity. The spatial accuracy
of the TOF modules is assumed to be a few mm.  
Both the barrel~\cite{TOF} and end-cap~\cite{CBM_TOF} TOF modules use similar
signal-amplification technology, however, the end-cap modules employ a modern, improved readout design. The barrel TOF modules were designed to
have a system resolution of 100 ps. This system has been demonstrated to have achieved a resolution 
of better then 90 ps~\cite{TOF}. The CBM TOF wall is designed to have a system time resolution of better than 80 ps. This is to be achieved with counters
having an intrinsic resolution of 60 ps and assuming that there will be a time-zero reference 
of better than 50~ps. This can be reached using the vertex position detector (VPD), which has been shown to provide a 
start time resolution of 24 ps in central $Au+Au$ collisions. The timing resolution will not be degraded through the angle of incidence as the range of angles in the $r-z$ plane will be from 70 to 50 degrees and in the $\phi-z$ plane the incident angle will always be greater than 45 degrees. As a conservative estimate, a time resolution of 100 ps for both the barrel and end-cap TOF arrays is assumed to generate the acceptance tables. A timing resolution performance of 80 ps for the system would extend the $p_T$ TOF PID limits by approximately 300 MeV/c for protons and 200 MeV/c for pions. 

 For mid-rapidity tracks with a 
flight path of 2.2~m, $\pi/K$ and $(\pi+K)/p$ separations are achieved for $p <$~1.6~GeV/c 
and 2.6~GeV/c, respectively. These separation cuts scale with an increase in
track length. The longest flight path for the barrel TOF are the $\eta$ = 0.96 tracks, which have
a path of 3.3~m. The eTOF is set back from the TPC end-cap at a distance of 2.8 m
from the interaction point. The longest flight paths for the eTOF are those at $\eta$ of 1.05,
which have paths of 3.6 m. The shortest paths (3.0 m) are for the tracks at $\eta$ = 1.7. 
The combined effects of timing, spatial, and momentum resolution are studied as a function of
$\eta$ and the TOF PID limits are shown as functions of $y$ and $p_T$ in 
Figs.~\ref{Acceptance_pro}, \ref{Acceptance_kap}, and \ref{Acceptance_pip}.

\begin{table}[h]
\caption{The flight path, radius of last TPC hit, momentum resolutions, and TOF PID limits for pions 
and protons at selected $\eta$.} 
\label{TOF_table}
\begin{tabular*}{0.5\textwidth}{@{}l*{15}{@{\extracolsep{0pt
          plus12pt}}l}}
$\eta$ & flight path & $r_{\rm last\,hit}$ & $\delta p/p (\pi)$ & $\pi/K$ & $\delta p/p$(pro) & $(\pi+K)/p$ \\
 & (cm) & (cm) & (\%) & (GeV/c) & (\%) & (GeV/c) \\
\hline
0.0 & 220 & 200 & 1.6  & 1.7 & 2.8 & 2.6 \\
\hline
0.5 & 248 & 200 & 1.6  & 1.6 & 2.7  & 2.4 \\
\hline
0.96 & 329 & 163 & 1.6  & 1.3 & 2.6  & 2.1 \\
\hline
1.05 & 358 & 160 & 1.6  & 1.3 & 2.5  & 2.1 \\
\hline
1.2 & 336 & 133 & 1.7  & 1.1 & 2.8  & 1.8 \\
\hline
1.3 & 325 & 118 & 1.8  & 1.0 & 2.9  & 1.6 \\
\hline
1.4 & 316 & 105 & 2.0  & 0.9 & 3.1  & 1.4 \\
\hline
1.5 & 309 & 94 & 2.2  & 0.8 & 3.4  & 1.3 \\
\hline
1.6 & 303 & 84 & 2.4  & 0.75 & 3.7  & 1.2 \\
\hline
1.7 & 299 & 76 & 2.7  & 0.7 & 4.1  & 1.1 \\
\hline
\end{tabular*}

\end{table}

\begin{figure}[th]
\includegraphics[scale=0.4]{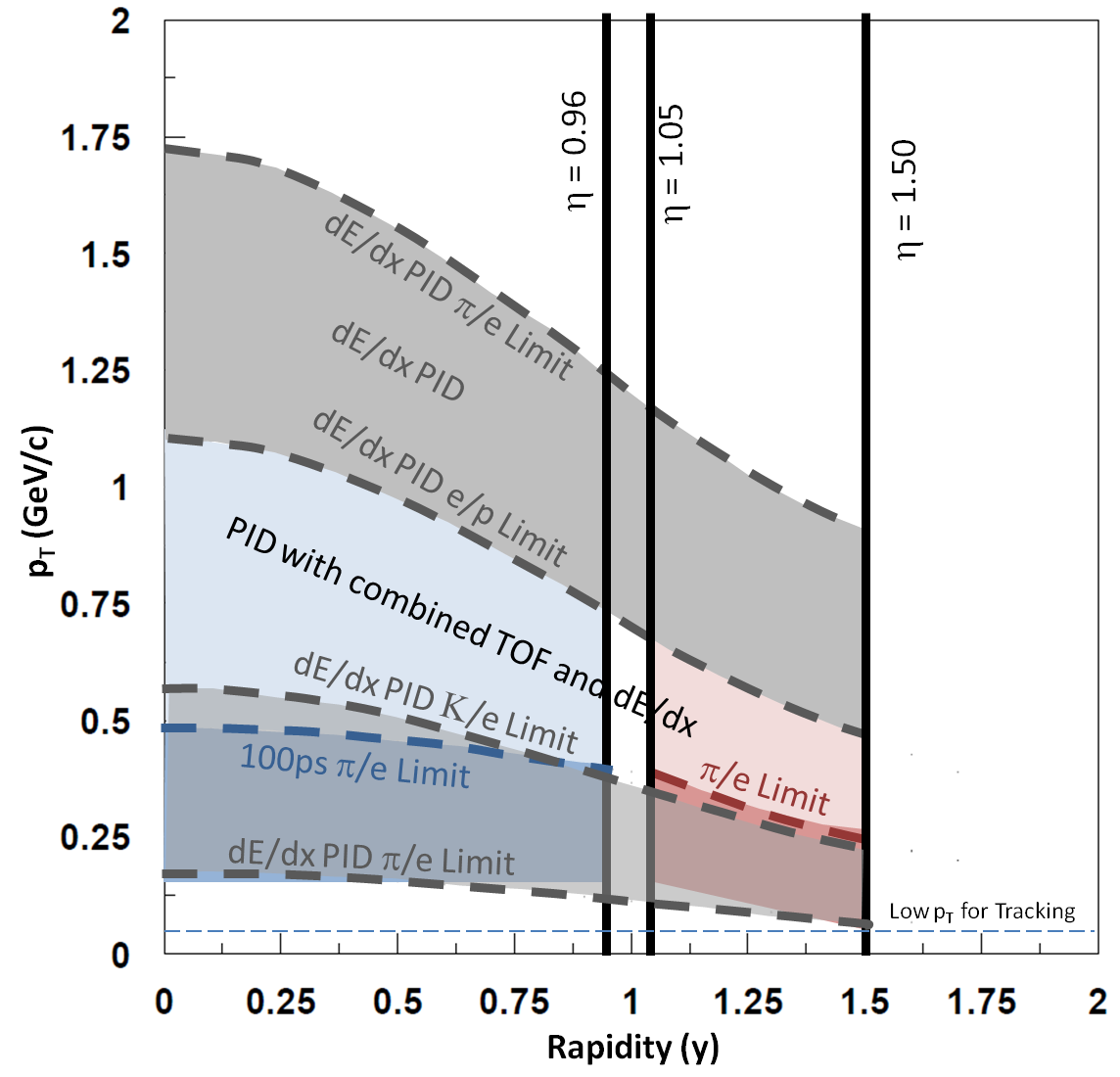}
\caption{The $p_T-y$ acceptance map for electrons showing the limits due to
tracking coverage and PID.}
\label{Acceptance_ele}
\end{figure}

It is also relevant to consider the acceptance and PID limitations for electrons. As 
all measured electrons are highly relativistic, pseudorapidity and rapidity are essentially
the same. The low-$p_T$ limit for electrons is set by the radius of curvature limits: 50 MeV/c 
tracking, 150~MeV/c to reach the barrel TOF, and $\eta$-dependent limits for the eTOF.
Electrons can be identified by TOF 
until they merge with the pions at 500 MeV/c. For much of this same $p_T$ range, the relativistic electrons also can be identified by $dE/dx$ between the pion and kaon bands which are falling
in the $1/\beta$ region of $dE/dx$-space. There is then a range of momenta from 0.5 to 1.1~GeV/c 
for which electrons cannot be cleanly identified by either TOF or $dE/dx$ alone. In this 
region, the electrons are merged with the pions in TOF-space  and with either kaons or protons 
in $dE/dx$-space. By using information from both systems, it is still possible to cleanly 
identify electrons. Finally, for momenta above 1.1 GeV/c, the 
relativistic electrons can be cleanly identified using $dE/dx$ up to 1.7 GeV/c when the pions 
approach their relativistic plateau. These acceptance regions are shown in Fig.~\ref{Acceptance_ele}.

All the acceptance windows shown in Figs.~\ref{Acceptance_pro}, \ref{Acceptance_kap}, 
\ref{Acceptance_pip}, and \ref{Acceptance_ele} were generated assuming that the 
collision was at $z$ = 0, in the center of the detector. However, for inclusive studies, the acceptance windows in rapidity can be affected by intentionally selecting events with a vertex offset as large as one meter because the longer bunches envisioned for BES-II will provide a broad interaction diamond. For offsets of $z$ = 1 m, the barrel TOF will extend from $\eta$ = -0.4 to 1.2 while the 
eTOF will cover from $\eta$ = 1.3 to 2.0. Selecting events with offset vertices will allow the 0.1 unit $\eta$ 
gap between the barrel and end-cap TOF systems to be covered.

\subsection{Rapidity Dependence of $p_T$ Spectra}
\begin{figure}
 \begin{center}
 \includegraphics[width=0.45\textwidth]{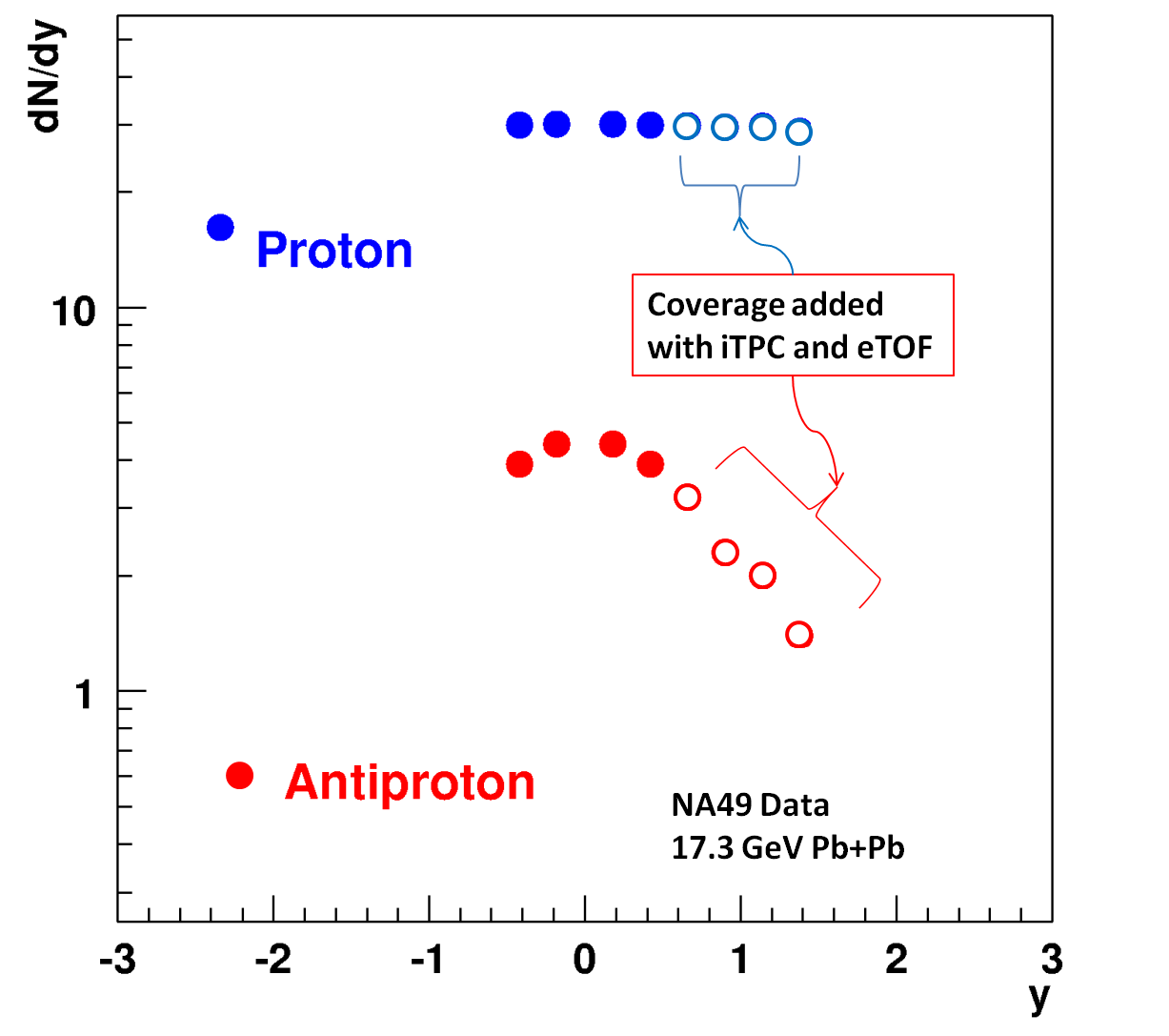}
 \caption{The $dN/dy$ values for protons from 17.3 GeV Pb+Pb data (circles) are 
shown~\cite{Alt:2006dk}. The 
closed symbols are within the coverage of the current configuration of STAR. The open symbols show
the extension of coverage which is enabled by the iTPC and eTOF upgrades.} 
 \label{Proton_dNdy}
 \end{center}
\end{figure}

At the top RHIC energies and at the LHC, there is a region of boost invariance 
at mid-rapidity. However, lower collision energies are characterized by incomplete transparency
and partial stopping. This is most readily observed by comparing the rapidity density
distributions of protons to those of anti-protons. Sample distributions are shown in 
Fig.~\ref{Proton_dNdy}~\cite{Alt:2006dk}. 
The anti-proton yield, which is entirely comprised of produced quarks,
can be well described by a Gaussian at mid-rapidity. The proton yield
is much flatter in $y$
and clearly the distribution is not a thermalized Gaussian. The anti-proton to proton ratio, which is 
the best indicator of the baryon chemical potential, changes dramatically as a function of
rapidity. For the data shown in Fig.~\ref{Proton_dNdy}, the change in the anti-proton to proton
ratio would suggest a change in $\mu_B$ of 50 MeV from $y=0$ to $y=1.2$. (Note the magnitude of the 
change depends on the collision energy). This change in the 
ratios also highlights why statistical equilibrium models extract quite different 
$T$ and $\mu_B$ values when using mid-rapidity versus full acceptance (4$\pi$) yield data. 
The figure highlights
why this added rapidity coverage with eTOF PID is so important for the BES-II program.
As the $\mu_B$ of the system is a function of the degree of stopping at a given energy and
centrality, it is important that this stopping be measured as directly as possible.  
Extended rapidity coverage allows for the study of bulk properties as a function of 
rapidity. The collision energy step size of the BES-II program was selected in order to 
measure $\mu_B$ steps of about 50-60 MeV. This is roughly the same change in $\mu_B$ expected 
when shifting from $y=0$ to $y=1.2$. We should expect to see similar changes in bulk properties
when shifting from one BES energy to the next as when shifting from mid to forward rapidity. 
For $y > 1.0$, the eTOF is required for PID, as seen in 
Figs.~\ref{Acceptance_pro}, \ref{Acceptance_kap}, and \ref{Acceptance_pip}.

Strange baryons and mesons allow one to carefully tease out the stopping of the quarks
from the participant nucleons. The $\Lambda$, with one $u$ and one $d$ quark, should
show two thirds of the stopping effects of the proton, while the $\Xi^-$, with only a single 
$d$ quark, should show effects at the one-third level. The $K^+$ carries an $u$ quark, and
should show a rapidity density that is broadened due to partial stopping,
while the $K^-$ carries a $\bar{d}$ and should therefore show no stopping effects.

\begin{figure}
 \begin{center}
 \includegraphics[width=0.45\textwidth]{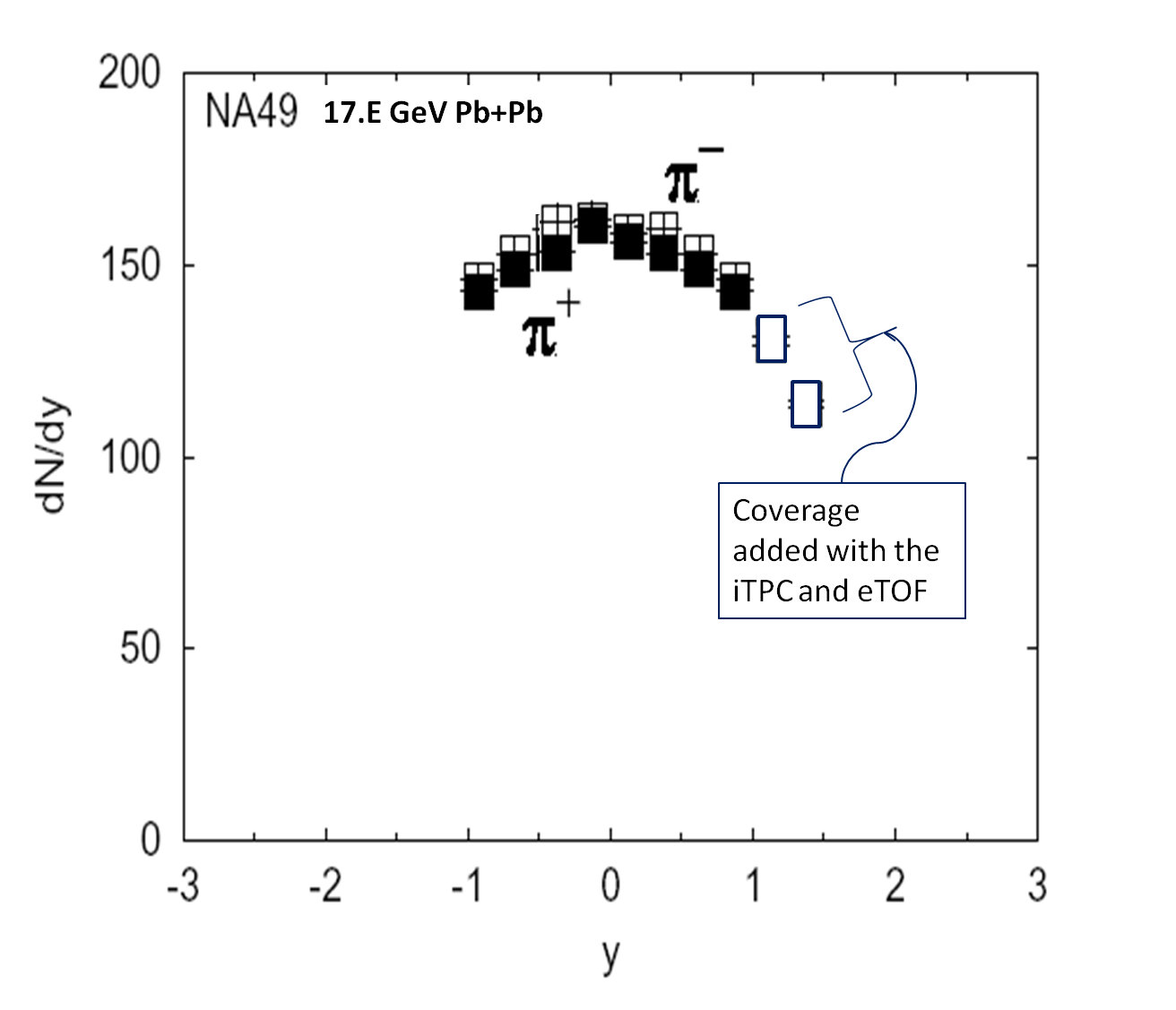}
 \caption{The $dN/dy$ values for pions from 17.3 GeV Pb+Pb data (squares) are 
shown~\cite{Chvala:2005rd}. The 
closed symbols are within the coverage of the current configuration. The open symbols show
the extension of coverage which is enabled by the iTPC and eTOF upgrades.} 
 \label{Pion_dNdy}
 \end{center}
\end{figure}

Pions are the most copiously produced particles. Although there are some isospin-dependent effects at 
the lowest center-of-mass energies and
at very low-$p_T$, the pions are for the most
part indicators of the freeze-out surface. The longitudinal extent of the pion rapidity
density distribution compared to the width suggested by Landau hydrodynamics has been
used as evidence for a drop in the speed of sound. This is indicative of a 
first-order phase transition~\cite{Petersen:2006mp,Alt:2007aa}. 
Determining the nature of the phase transition as a
function of collision energy is one of the key physics goals of the BES-II program.  
Studying the widths of the pion rapidity distributions provides evidence of
the expected softening of the equation of state. 
The capability of the STAR detector to measure the
pion rapidity density width is illustrated in Fig.~\ref{Pion_dNdy}. Data from
NA49 for Pb+Pb collisions is shown~\cite{Chvala:2005rd} 
in the acceptance window of the current configuration
(solid symbols) and with the extended rapidity and PID of the eTOF upgrade (open symbols).
In order to determine accurately the width of a Gaussian, the measurement window should 
be broader than one $\sigma$. For the energy range of the BES-II program, the 
pion rapidity widths are expected to range from 1.1 to 1.6 units of rapidity as the 
collision energy increases from 7.7 to 19.6 GeV~\cite{Rustamov:2012np}.  

\subsection{Dileptons}
Short-lived vector mesons, which decay into $e^+e^-$ (dilepton) pairs  
are seen as probes of the earliest stage of a 
heavy-ion reaction, because the daughter electrons escape the colored 
medium without substantial final-state interactions. The transition from a QGP to a dense hadron gas involves 
not only a deconfinement transition but also a spontaneous breaking of chiral  
symmetry. Chiral symmetry predicts that the spectral functions of chiral partners 
($\rho$ and $a_1$ for example) become degenerate in the symmetric phase. 
Since it is impossible in heavy-ion collisions to measure a spectral function for the 
$a_1(1260)$ meson, one cannot directly observe the disappearance of the mass splitting 
between the $\rho$ and $a_1(1260)$ experimentally. Instead, efforts are devoted to 
studying the modification of vector meson spectral functions in a hot dense medium.
 
A similar broadening of the mass of the $\rho$ has been observed from the top SPS energy~\cite{Adamova:2006nu, Adamczyk:2015mmx} to the top RHIC energy. 
This broadening causes an excess in the 
low-mass region (200 to 770 MeV/$c^2$) of the dilepton invariant mass 
spectrum. Using the broadened $\rho$ spectral function, QCD and Weinberg sum rules, and inputs from Lattice QCD, theorists have demonstrated that when the temperature reaches 170 MeV, the derived $a_1(1260)$ 
spectral function is the same as the $\rho$ spectral function. This is a signature of chiral 
symmetry restoration.
In a model calculation which describes the experimental data, the coupling to the baryons in the medium 
plays a dominant role in the broadening of the $\rho$ spectral function.
The ratio $(p + \bar{p})/(\pi^+ + \pi^-)$, which 
is a proxy for the total baryon density, remains fairly constant at mid-rapidity  
from top RHIC energies down to the top SPS energy, and then increases as one 
goes down through the BES-II range~\cite{Geurts:2013}. This behavior predicts a change in the normalized low-mass dilepton 
excess of a factor of two over the collision energies of 7.7 to 19.6 GeV. 
As can be seen in Figs.~\ref{Proton_dNdy} and \ref{Pion_dNdy}, one can also change  
the $(p + \bar{p})/(\pi^+ + \pi^-)$ ratio by shifting the analysis 
frame from mid-rapidity to forward rapidity. This rapidity dependence will 
provide a strong and independent observable to study the total baryon density dependence of the 
low-mass dielectron emission. Knowing the mechanism that causes in-medium $\rho$ broadening and the 
temperature and baryon density dependence is fundamental to our understanding and assessment of chiral symmetry restoration in hot QCD matter.
 
Due to the high hadron background, the quality of the PID is typically  
the primary limitation for dielectron measurements. Even with the iTPC upgrade,  
the electron identification would still be limited to the pseudorapidity range between  
$\pm1$. Electrons are always in the relativistic rise region of $dE/dx$ for gas 
ionization chambers. Therefore, clean PID requires another discriminating  
measurement such as the time of flight. With the eTOF upgrade, we can 
extend the electron identification to the range $|\eta|<1.5$. 
The STAR detector during BES-II
will have a unique capability to quantify the total baryon density
effect on the $\rho$ broadening. The improved measurements during BES-II
will enable us to distinguish models with different $\rho$ broadening
mechanisms, for example, the Parton-Hadron String Dynamic (PHSD)
transport model~\cite{Linnyk:2011hz,Linnyk:2011vx} versus a microscopic many-body model with
macroscopic medium evolution~\cite{Rapp:2000pe,vanHees:2006ng}. The rapidity-dependent measurements during
BES-II enabled by the eTOF will provide complementary information on
this important physics topic.
 
\subsection{Directed Flow}
Proton directed flow ($v_1$) measurements from the BES-I program have shown 
a very intriguing behavior~\cite{Adamczyk:2014ipa}; no model yet shows good agreement, 
and the favored interpretations by theorists include both a crossover and a first-order 
phase transition~\cite{Stoecker:2004qu, PhysRevC.89.054913, PhysRevC.90.014903, phsd2, 
Nara:2016phs}. 
The mid-rapidity slope 
$dv_1/dy$ switches from positive to negative between $\sqrt{s_{NN}} = 7.7$ and 11.5 
GeV, and reaches a minimum near 14.5 GeV. The slope $dv_1/dy$ for net-protons has a 
similar minimum but then switches back to a positive slope between 27 and 39 GeV.  
This behavior could indicate a repulsive compression at the lowest and highest energies, and a 
softening of the equation of state, consistent with a spinodal decomposition, at the 
intervening beam energies.  Even though this remarkable result still needs theoretical 
progress to provide interpretation, further experimental tests can help elucidate 
the underlying physics.  

During the evolution of a heavy-ion collision, gradients of 
pressure, density, and temperature are established across the interaction zone. 
The lateral edges of the collision will have lower pressure and will be shifted in 
rapidity in the direction of the adjacent spectator matter. Thus while we might 
achieve spinodal decomposition in the center of the collision zone at a particular beam 
energy, the edge regions might still undergo repulsive compression due to the shifts 
forward and backward in rapidity. This would in turn affect the $v_1$ slope for 
protons as a function of rapidity --- the so-called wiggle.  While the 
mechanism mentioned above might not be adequate to explain the wiggle phenomenon in its entirety, it is 
plausible to expect it to modify the wiggle phenomenology. Therefore, a comprehensive 
mapping of the $v_1(y)$ structure at BES energies will offer new insights into key 
details of the QCD equation of state in the relevant region of the phase diagram.  Although NA49 
reported some evidence along these lines~\cite{Alt:2003ab}, 
a more comprehensive study is needed for conclusive results. 
 
\begin{figure} 
 \begin{center} 
 \includegraphics[width=0.5\textwidth]{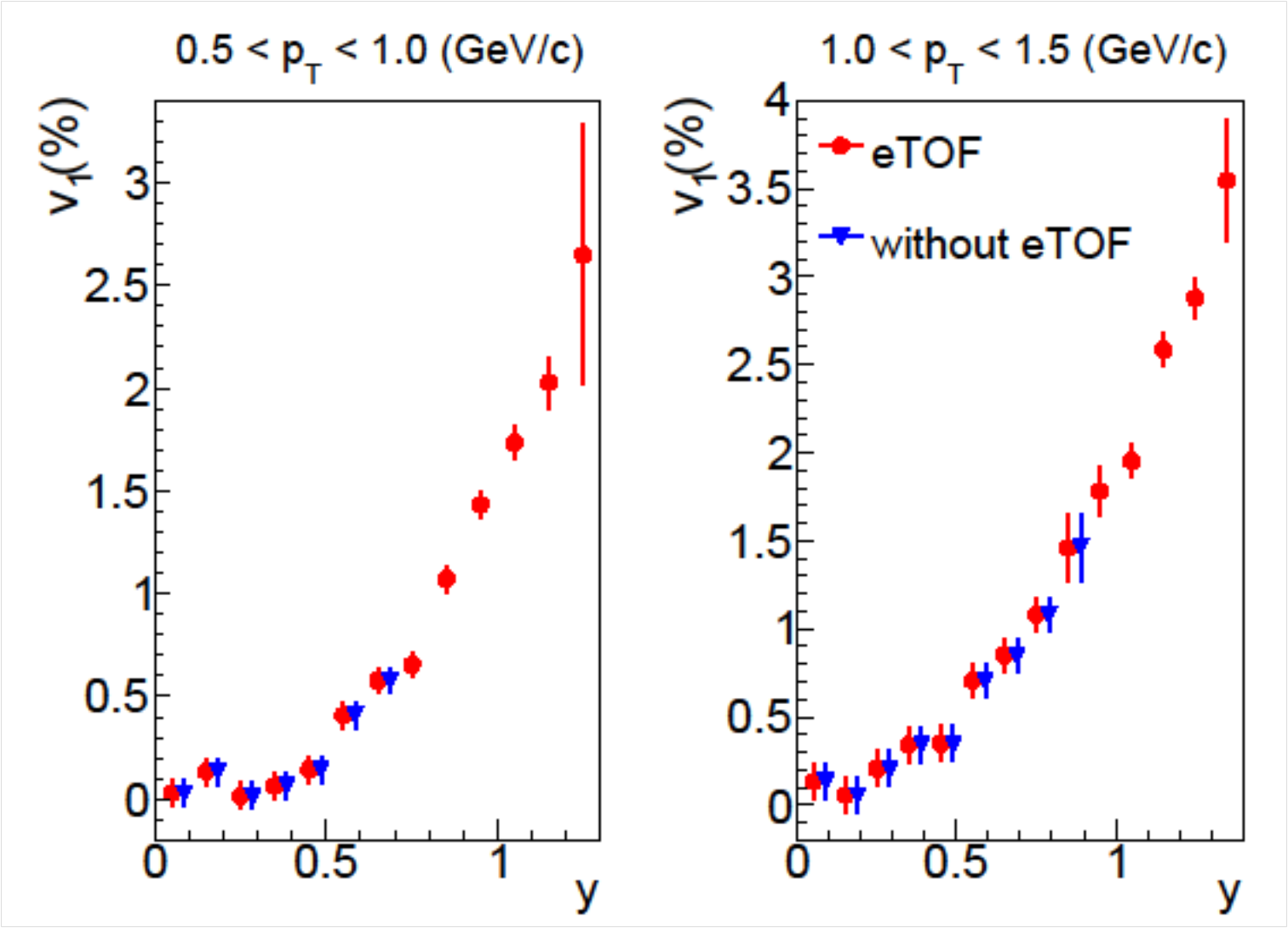} 
 \caption{Proton directed flow as a function of rapidity for minimum-bias Au + Au 
collisions at $\sqrt{s_{NN}} = 19.6$ GeV, based on the UrQMD model. The simulated $v_1(y)$ in two
intervals of $p_T$ is compared between the acceptance of the
STAR iTPC with the existing TOF barrel (blue triangles) and
the upgraded acceptance after addition of the eTOF (red circles). }  
 \label{Proton_v1_UrQMD} 
 \end{center} 
\end{figure}

The eTOF will provide proton identification up to a rapidity of 1.2, enabling a 
study of $v_1(y)$ over a new rapidity region for protons, kaons, and pions.  Figure~\ref{Proton_v1_UrQMD}, based on protons from 0.9 million minimum-bias UrQMD model events at 
$\sqrt{s_{NN}} = 19.6$ GeV, illustrates the new parameter space opened up by 
the eTOF.  Figure~\ref{Proton_v1_UrQMD} 
assumes ideal reaction plane resolution, and the plotted 
statistical errors are typically a couple of times larger than the expected 
statistical errors for 7.7 to 19.6 GeV BES-II running using fine centrality 
bins. The $v_1(p_T)$ for 
two different $p_T$ intervals 
is shown in the panels of the figure. The $p_T$ dependence of every 
$v_n$ Fourier coefficient is, {\it a priori}, of empirical interest. A good illustration 
of this is provided by constituent quark scaling and its role in QGP discovery as 
originally revealed by measurements of $v_2(p_T)$ for mesons and baryons. It is evident from Fig.~\ref{Proton_v1_UrQMD} that the steepening of the proton $v_1(y)$ slope 
beyond the mid-rapidity region in the relevant regions of $p_T$ is poorly 
resolved without eTOF, and a substantial region of new phenomenological 
coverage is opened up by eTOF.

\subsection{Elliptic Flow}
\begin{figure}
 \begin{center}
 \includegraphics[width=0.45\textwidth]{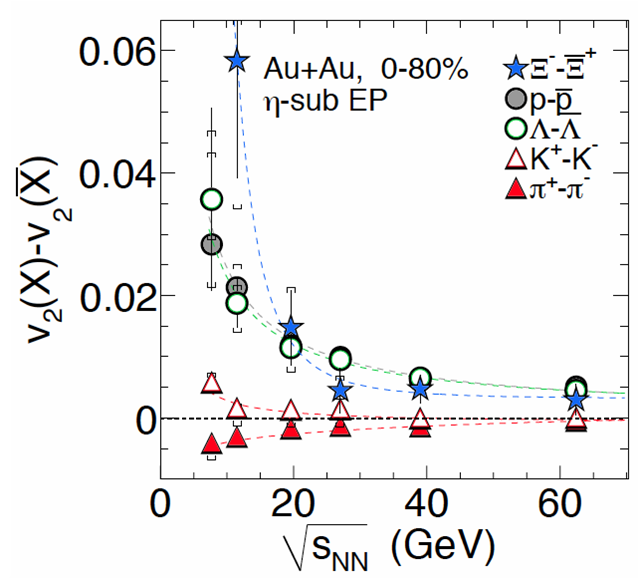}
 \caption{The measured difference in integrated $v_2$ between particles and their corresponding
antiparticles: pions (filled triangles), kaons (open triangles), $\Lambda$s (open circles), protons (filled circles), and $\Xi$s (filled stars)~\cite{Adamczyk:2013gv}.} 
 \label{ParticleAntiparticle_v2}
 \end{center}
\end{figure}

Number of constituent quark scaling (NCQ) of elliptic flow has been seen as one of the cornerstone 
pieces of evidence that collectivity is established on the partonic level at the  
top energy of RHIC~\cite{Adams:2004wz}.
One of the goals of the BES program is to see how these key
QGP signatures evolve with collision energy. 
Although the quark number scaling of  
elliptic flow seems to hold qualitatively for particles and for  
anti-particles above 19.6~GeV~\cite{Adamczyk:2013gw} 
(the statistics are limited below 19.6 GeV), 
when one compares the $v_2$ of particles to their respective anti-particles one
sees a very different trend. This is shown in 
Fig.~\ref{ParticleAntiparticle_v2}~\cite{Adamczyk:2013gv}.
This discrepancy could be suggesting a break down in the scaling behavior. It
could also be indicating a more subtle effect coming from the incomplete transparency and 
partial stopping of the valence quarks. A 
possible explanation for this behavior is that transported quarks have a very 
different flow profile from quarks created in the fireball~\cite{Dunlop:2011cf}. 
This conjecture could 
be tested by studying elliptic flow at more forward rapidity where the ratio
of transported quarks to created quarks is much higher than at mid-rapidity. The particle to 
anti-particle $v_2$ differences are expected to increase 
significantly for $y>1.0$. The eTOF will enable these rapidity-dependent measurements
of $v_2$. This can help us better understand the nature of this QGP signal and whether it
disappears or is simply obscured by other effects as the
collision energy is reduced. 
It must be 
demonstrated that the changes in the signature with energy are an effect of QGP physics.

The $\phi$ is a particularly interesting case because it is a meson with a
mass close to that of a nucleon. Determining the constituent quark flow behavior of the $\phi$ 
would be a sensitive test of whether flow is established at the
partonic level, especially because there is no transported valence 
quark effect. The results for the flow of the $\phi$ at the lowest energies
of the BES-I are suggestive but far from conclusive  due to marginal statistics. This open 
question may be answered in the BES-II program. Increased
luminosities are expected to be provided by extending the bunch length at all
BES-II energies and by electron cooling at the lowest three energies 
(7.7, 9.1, and 11.5 GeV). However, even with these increased luminosities, the $v_2$ of the
$\phi$ is still one of the most statistically demanding measurements proposed
for BES-II~\cite{BESII}. Since this is one of the top statistics drivers of the program, 
any upgrade 
that improves acceptance for the $\phi$ directly improves the
program. The $\phi$ is detected through the decay to a $K^+ K^-$ pair. 
The iTPC improves the kaon acceptance at low-$p_T$.
The eTOF provides kaon identification up to 1.6 GeV/c in the
extended pseudorapidity range $|\eta|<1.5$. This would lead to a more significant 
$\phi$ signal.

\subsection{Fluctuations - Higher Moments of Conserved Quantities}
\begin{figure}
 \begin{center}
 \includegraphics[width=0.45\textwidth]{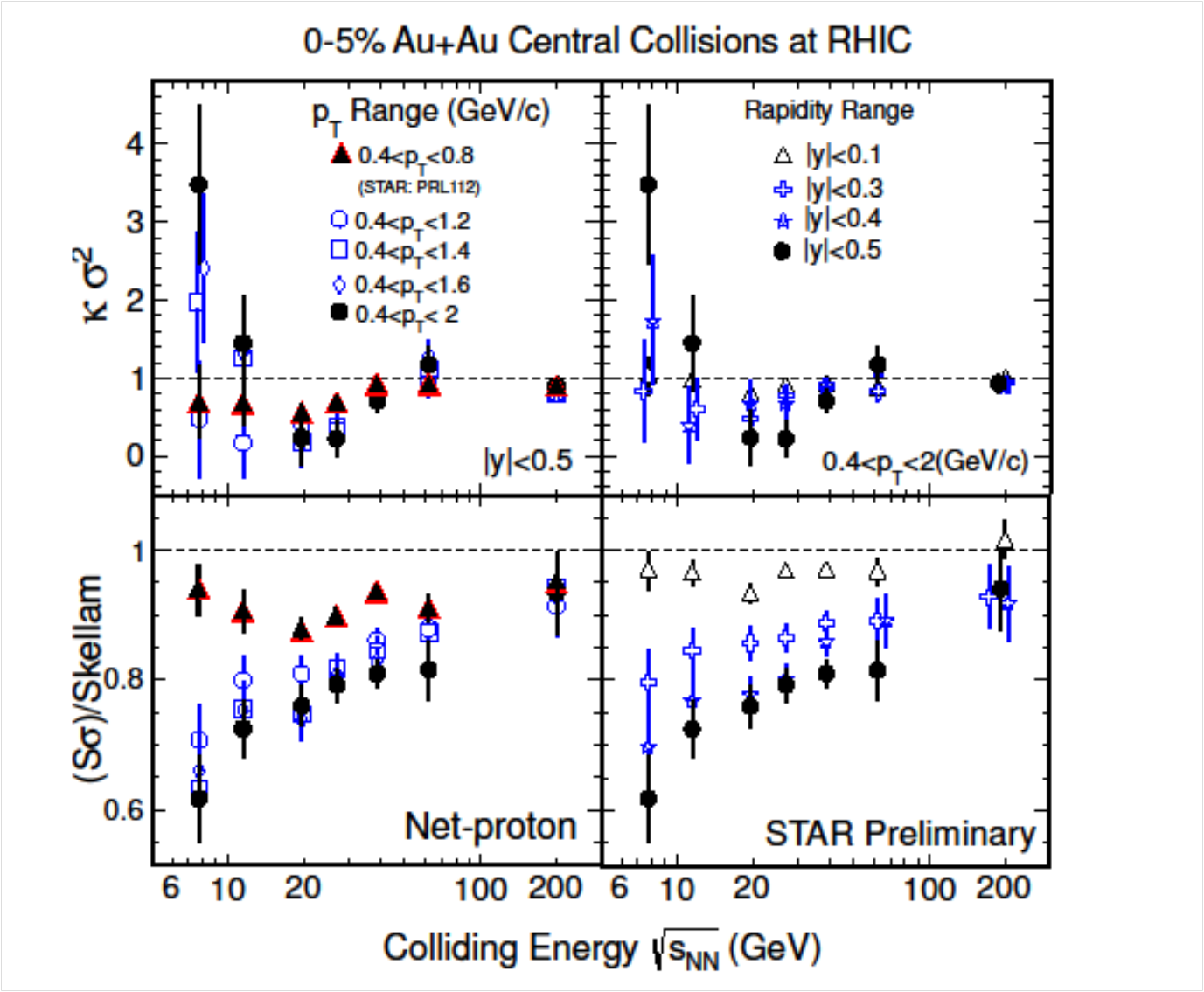}
 \caption{STAR results for beam energy dependence of $\kappa\sigma^2$ (top panels) and 
$S\sigma$/Skellam (lower panels) for net-protons in Au+Au 
collisions~\cite{Adamczyk:2013dal}. The left panels 
illustrate the effect of $p_T$ selection while the right panels indicate the effects 
of rapidity selection. Dotted horizontal lines are expectations from Poisson
distributions.} 
 \label{Kurtosis}
 \end{center}
\end{figure}

\begin{figure}
 \begin{center}
 \includegraphics[width=0.45\textwidth]{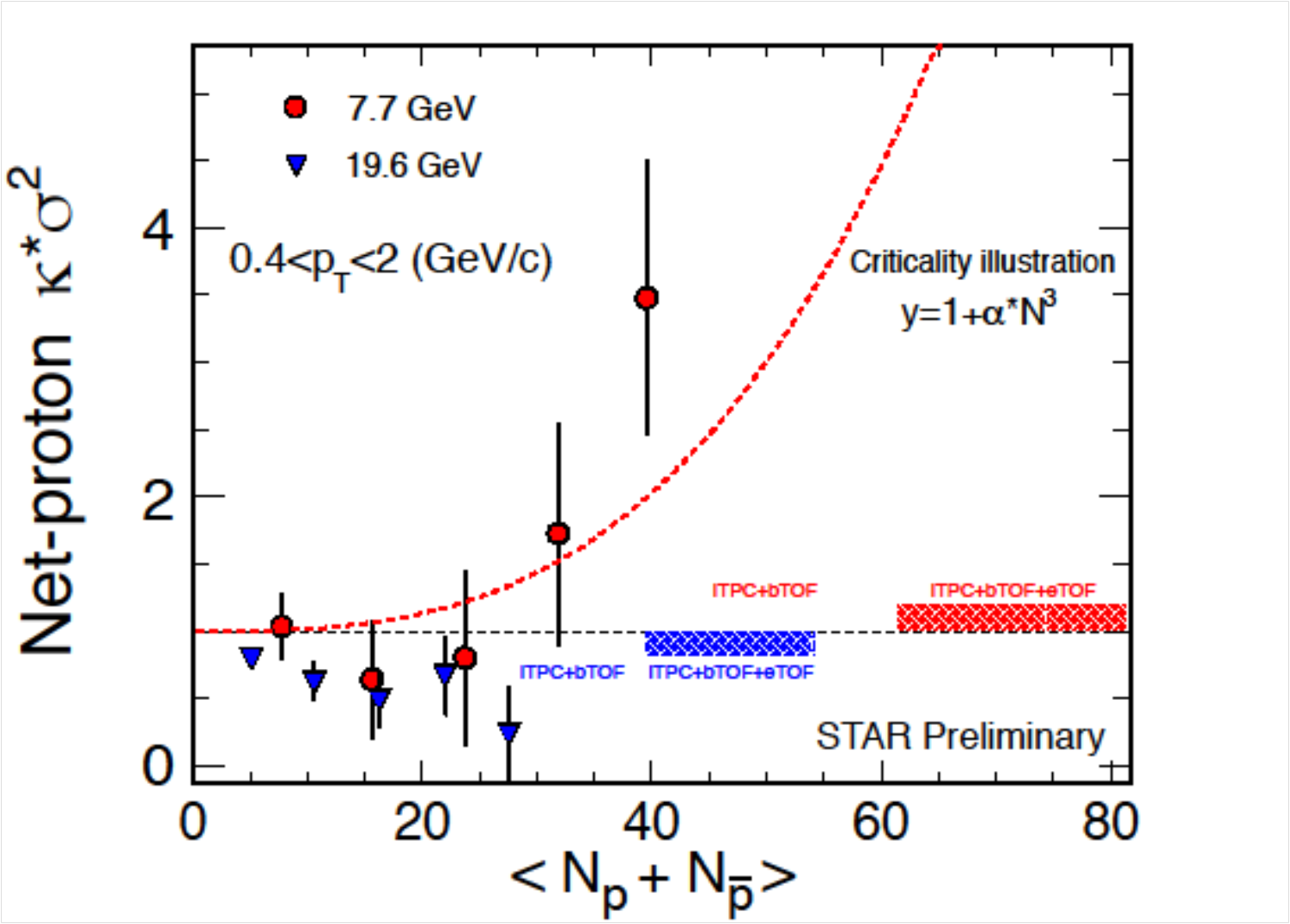}
 \caption{The net-proton $\kappa \sigma^2$ as a function of 
the protons and anti-proton multiplicity. Protons come from the rapidity range -1.4 $< y_P <$ 0.9.} 
 \label{Kurtosis_etof}
 \end{center}
\end{figure}

Measurements of net-proton (proxy for net-baryon) and net-kaon (proxy for net-strangeness) kurtosis times variance 
($\kappa\sigma^2$), which is the same as the cumulant ratio $C_4/C_2$, are 
likely the best indicators of 
critical behavior in the vicinity of a possible critical point in the QCD phase diagram. 
We have observed that the net-proton fluctuation signals strongly depend on the $p_T$
and rapidity cuts of the protons (see Fig.~\ref{Kurtosis}).
The net-proton fluctuation analyses have used cuts of $0.4 < p_T < 2.0$ GeV/$c$. Using the
current TPC, the rapidity is cut at $\pm0.5$ ($\Delta y = 1.0$), while with the iTPC, 
this cut can be extended to $\pm0.8$ ($\Delta y = 1.6$). Additional particle identification from 
the eTOF extends the rapidity reach. However as the rapidity is extended past 0.8, 
the hard $\eta=1.5$ acceptance cut imposes a varying low-$p_T$ cut. This requires a different analysis approach. Instead of presenting the sensitivity of $\kappa\sigma^2$ as a function of the width of the rapidity window as was done in Fig.~\ref{Kurtosis}, 
we use $\eta$ cuts, which can be opened symmetrically, and consider the
sensitivity as a function of the sum of the number of measured protons and 
anti-protons. 
This analytical technique is shown in Fig.~\ref{Kurtosis_etof}. The STAR BES-I 
data for 7.7 GeV trend upward with total baryons while for 19.6~GeV, the trend is downward. 
It is expected that the $\kappa\sigma^2$ signal will be large for energies that create systems near the critical point, while for systems with a baryon chemical potential below the critical point, the $\kappa\sigma^2$ will drop 
below unity. The added coverage of the eTOF will enhance the fluctuation signal providing a clearer 
and more significant indication of critical behavior.

The addition of the eTOF for PID will have a significant impact on the net-kaon 
and net-charge (which is directly measured from the yields of positive and negative hadrons) fluctuation 
analyses. The eTOF will allow an extension 
of the analysis window for net-kaons to $y = 1.2$ and for net-charge to $\eta = 1.5$. 

\section{Extending the Energy Scan Below 7.7 GeV with an Internal Fixed-Target and eTOF}
\begin{figure}[th]
\includegraphics[scale=0.4]{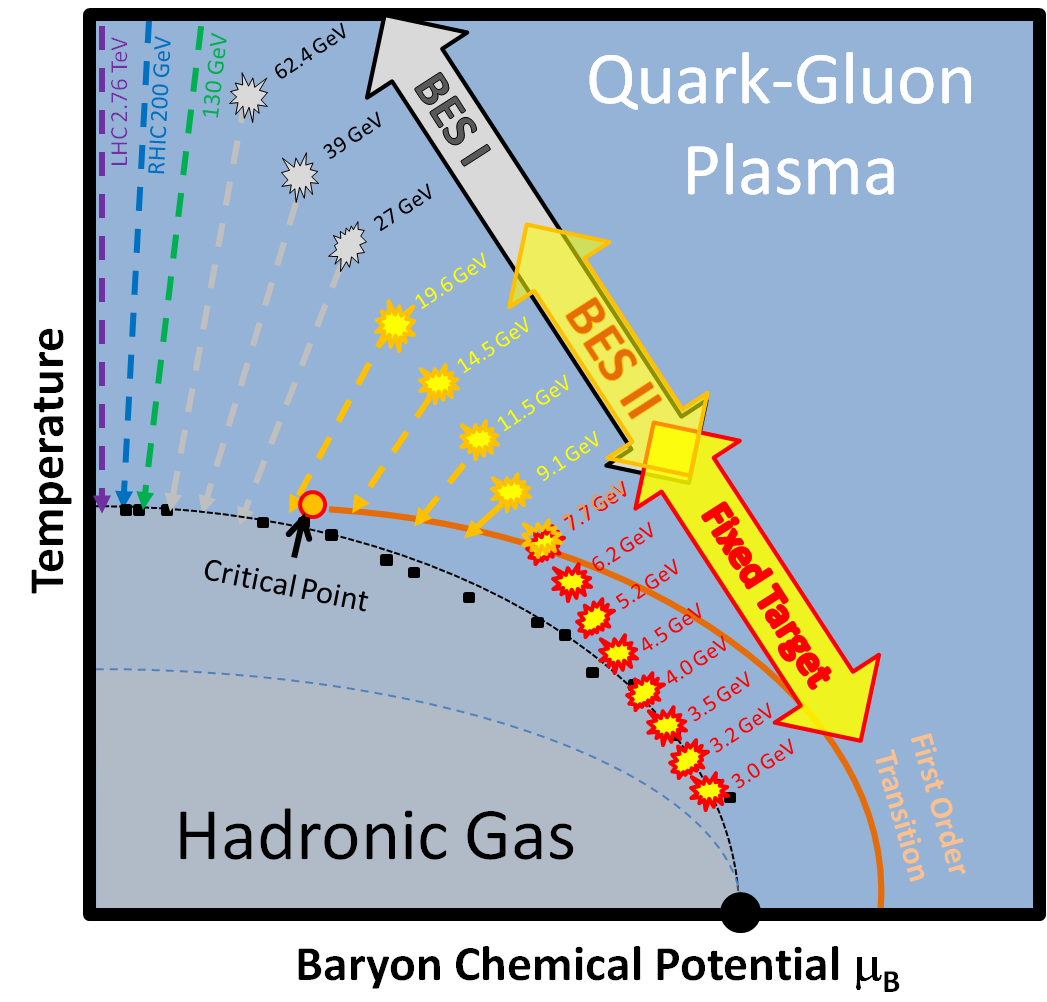}
\caption{A schematic of the phase diagram of QCD matter showing hypothetical illustrations
of reaction trajectories for the BES collider and fixed-target programs.}
\label{PhaseDiagramFXT}
\end{figure}

In normal collider mode, the lowest collision energy available
at RHIC is 7.7 GeV. The collider tried circulating beams at 5.0 GeV
but because the luminosity is proportional to $\gamma^3$ (the relativistic
$\gamma$ of the individual ion beams), operating below 7.7 GeV proved
to be impractical. It is important to measure key
observables at energies lower than 7.7 GeV for several reasons:
\begin{itemize}
\item{NA49 has reported that the onset of deconfinement occurs at 
7.7 GeV~\cite{Alt:2007aa}.}
\item{Some of the QGP signatures (local parity violation~\cite{Adamczyk:2014mzf}
and narrowing of balance functions~\cite{BES_Balance_Functions}) show signs of 
disappearing at 7.7 GeV. We need to extend the energy range so that we can 
confirm that these signatures have indeed turned off.}
\item{There are theoretical calculations suggesting that the mixed phase is entered at 
energies well below 7.7 GeV~\cite{Steinheimer:2012gc}.}
\end{itemize} 
The fixed-target program at STAR with the eTOF upgrade will enable the energy scan to 
extend below 7.7~GeV and address these questions.
With the eTOF, the fixed-target program can study the center-of-mass energy region from $\sqrt{s_{NN}}$ = 3.0 to 7.7 GeV.  
Some of the AGS, SPS, and SIS collaborations use projectile kinetic energy per nucleon; in that notation,
the above range corresponds to 2.9 to 30.3 AGeV. 
The five energies of the BES-II collider program cover the baryon chemical potential 
range from 205 to 420 MeV~\cite{Cleymans:2005xv}. The inclusion of an additional seven 
fixed-target energies will extend the range from 420 to
720 MeV with a similar 50 MeV step size (see Fig.~\ref{PhaseDiagramFXT}).
The physics topics proposed for normal 
collider mode can be performed in this extended $\mu_B$ range.

It is important to note that the physics impact of the eTOF system is significantly 
different for the fixed-target program than for collider mode. In the BES-II collider 
program, the addition of the eTOF system extends the momentum range of forward rapidity PID. 
In the fixed-target program, the roles of the central and forward parts of the detector 
are reversed. This is due to the large 1-2 unit rapidity offsets between the laboratory 
and the center-of-mass reference frames. In the next section, we show that eTOF 
provides essential PID in the mid-rapidity region.

\subsection{Acceptance}
\begin{figure}[th]
\includegraphics[scale=0.4]{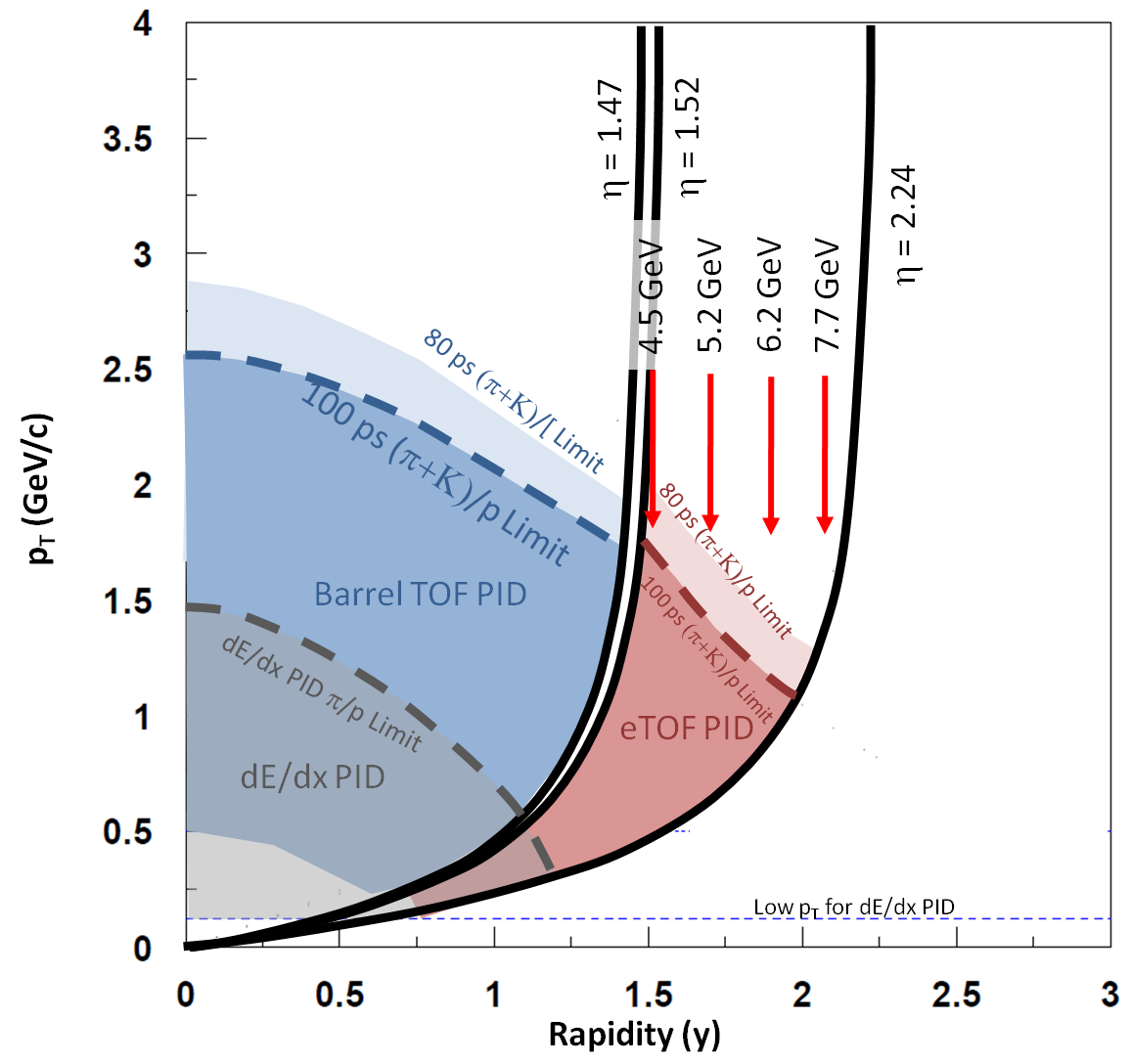}
\caption{The $p_T-y_{lab}$ acceptance map for protons in the fixed-target configuration showing the limits due to tracking coverage and PID. Arrows indicate $y_{CM}=0$ for
the $\sqrt{s_{NN}}$ = 4.5, 5.2, 6.2, and 7.7 GeV energies.}
\label{Acceptance_FXT_pro}
\end{figure}

\begin{figure}[th]
\includegraphics[scale=0.4]{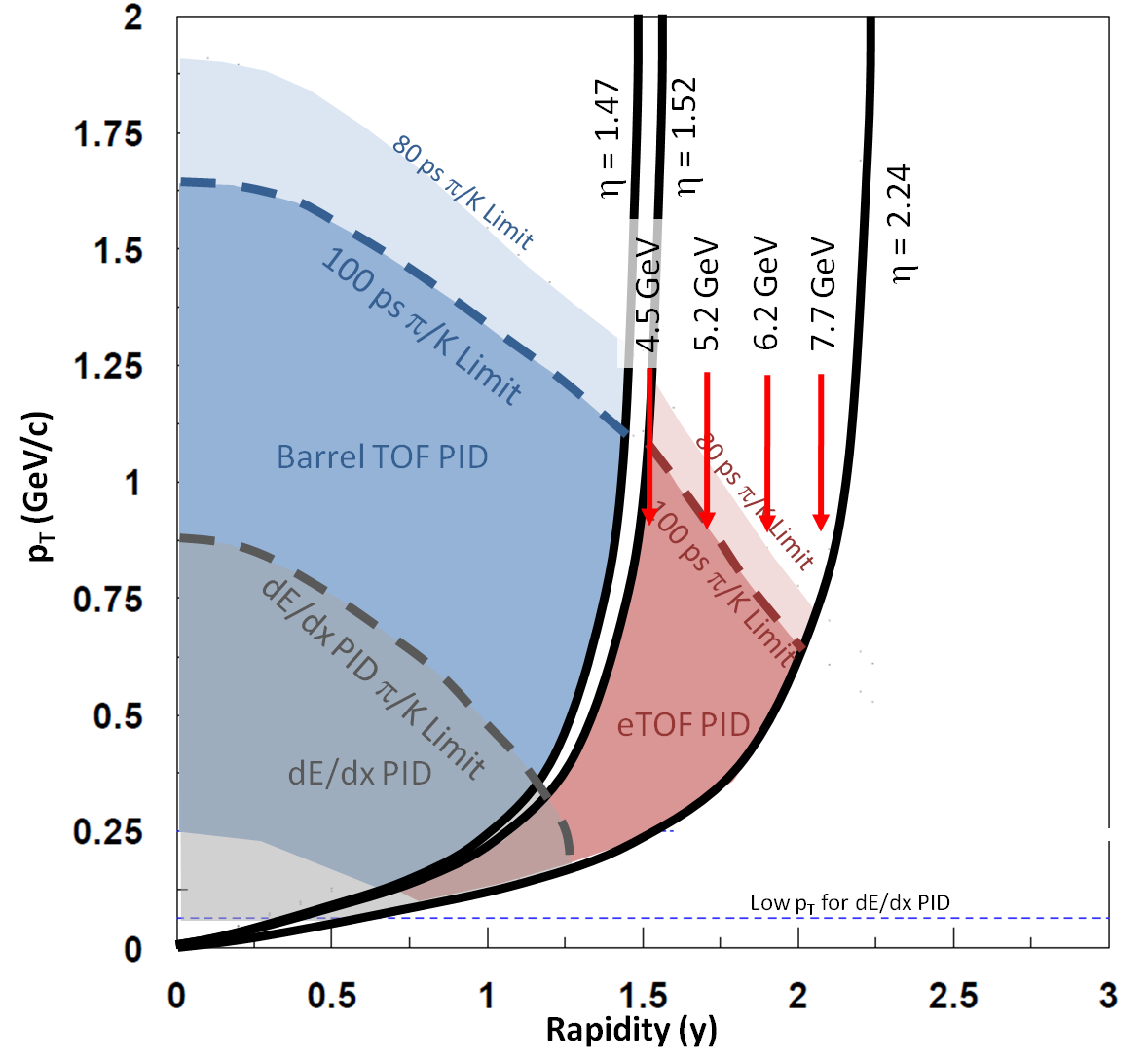}
\caption{The $p_T-y_{lab}$ acceptance map for kaons in the fixed-target configuration showing the limits due to tracking coverage and PID. Arrows indicate $y_{CM}=0$ for
the $\sqrt{s_{NN}}$ = 4.5, 5.2, 6.2, and 7.7 GeV energies.}
\label{Acceptance_FXT_kap}
\end{figure}

\begin{figure}[th]
\includegraphics[scale=0.4]{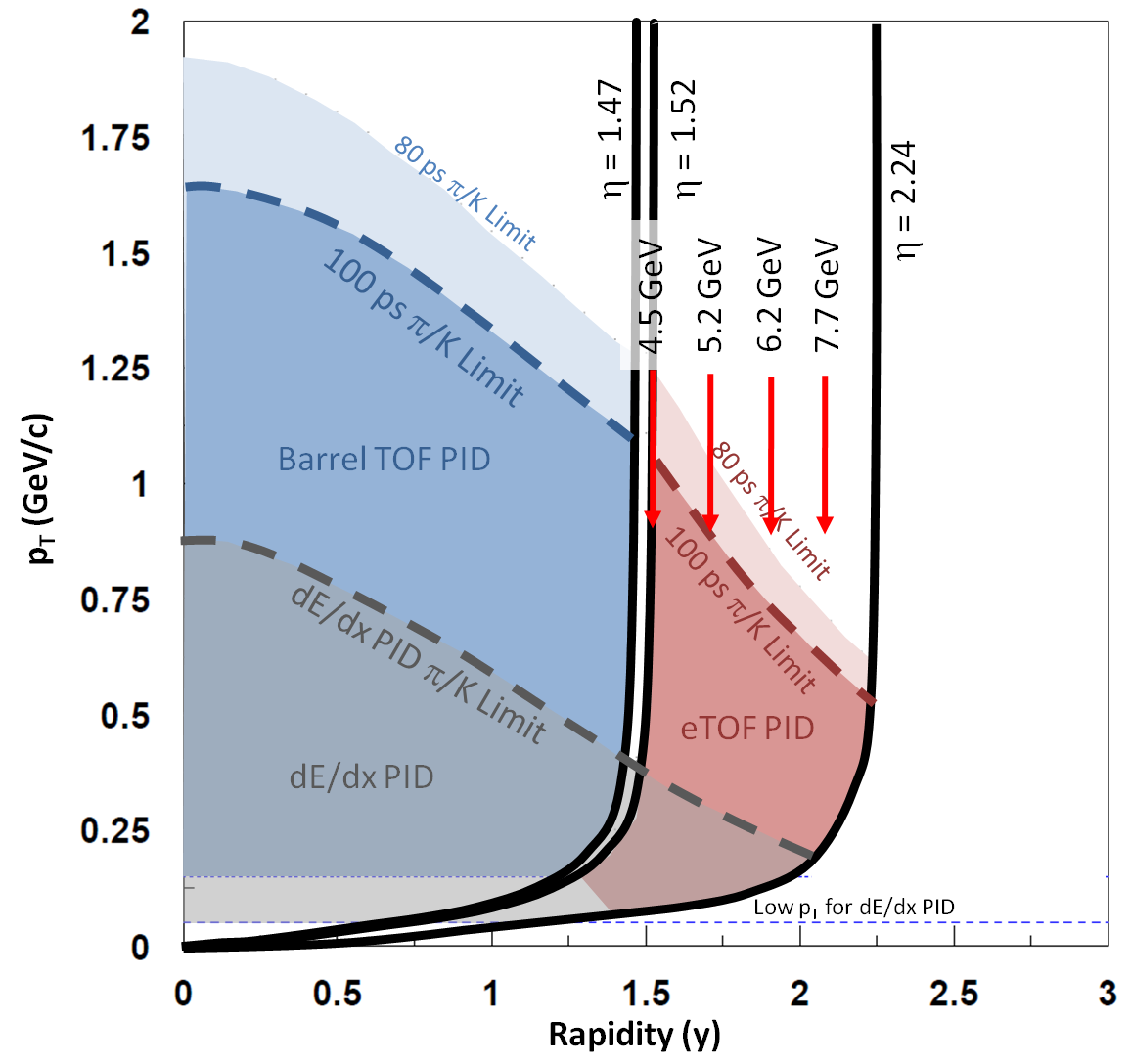}
\caption{The $p_T-y_{lab}$ acceptance map for pions in the fixed-target configuration showing the limits due to tracking coverage and PID. Arrows indicate $y_{CM}=0$ for
the $\sqrt{s_{NN}}$ = 4.5, 5.2, 6.2, and 7.7 GeV energies.}
\label{Acceptance_FXT_pip}
\end{figure}

\begin{figure}[th]
\includegraphics[scale=0.4]{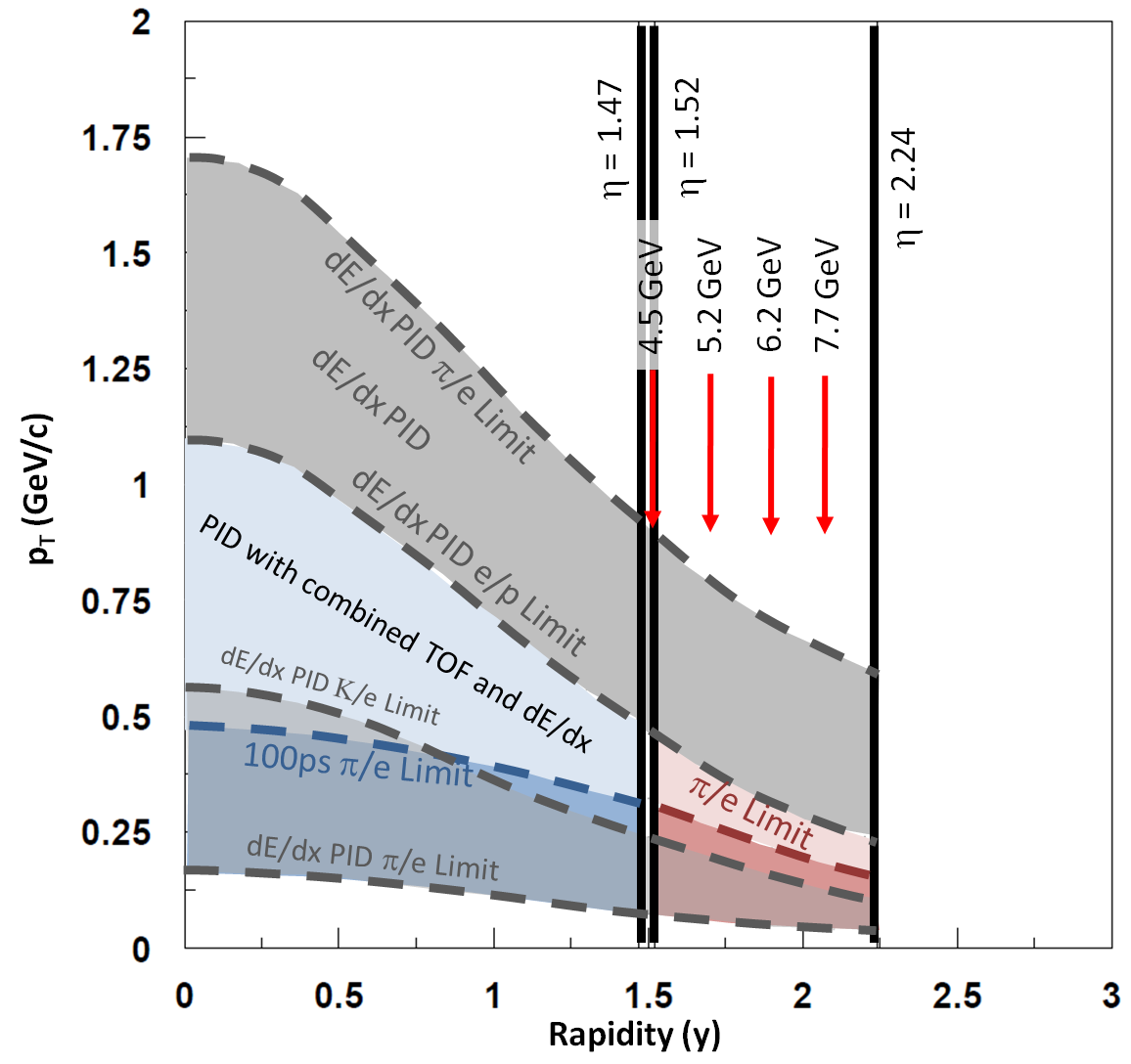}
\caption{The $p_T-y_{lab}$ acceptance map for electrons in the fixed-target configuration 
showing the limits due to tracking coverage and PID. Arrows indicate $y_{CM}=0$ for
the $\sqrt{s_{NN}}$ = 4.5, 5.2, 6.2, and 7.7 GeV energies.}
\label{Acceptance_FXT_ele}
\end{figure}

 The calculation of the fixed-target acceptance of the STAR detector is similar to the collider mode acceptance calculations discussed in the previous 
section with only a few exceptions. 
The 1 mm thick gold target is located at $z = $ +210 cm. This is 
the optimal location for
the target because it allows measurements from target rapidity to mid-rapidity. The 210 cm shift in the 
location of the interactions has the following effect on the the acceptance and PID limits of STAR:
\begin{itemize}
\item{The low-$p_T$ threshold values for the eTOF are defined by energy loss in the backplane
of the TPC, and these remain unchanged.} 
\item{The $\eta$ limits of the detector are changed. The barrel TOF system covers up to $\eta$ = 1.47. The eTOF system covers $\eta$ = 1.52 to 2.24.} 
\item{The track length in the STAR TPC for particles with $\eta > 0.88$
is longer in fixed-target events. Therefore, 
the $dE/dx$ resolution for these tracks is better than for tracks 
with similar $\eta$ values in collider events. A sampling of the $dE/dx$ resolution is
given in Table~\ref{FXT_dEdx_table}.}
\item{The flight path for particles with $\eta > 0.96$
is longer in fixed-target events. Therefore, 
the TOF PID limits for these tracks extend to higher momentum than for 
tracks with similar $\eta$ in collider events. A sampling of the PID limits using TOF is
given in Table~\ref{FXT_TOF_table}.}
\end{itemize}
The acceptance and PID ranges for fixed-target events are shown in 
Figs.~\ref{Acceptance_FXT_pro}, \ref{Acceptance_FXT_kap},
\ref{Acceptance_FXT_pip}, and \ref{Acceptance_FXT_ele}.

\begin{table}[h]
\caption{The track lengths, $dE/dx$ resolutions, and $dE/dx$ PID limits for
various values of $\eta$ for fixed-target events.}
\label{FXT_dEdx_table}
\begin{tabular*}{0.5\textwidth}{@{}l*{15}{@{\extracolsep{0pt
          plus12pt}}l}}
$\eta$ & Track Length & $\sigma_{dE/dx}$ & $\pi/K$ & $p/\pi$ \\
 & (cm) & \% & (GeV/c) & (GeV/c) \\
\hline
0.0 & 126 & 5.5 & 0.88 & 1.47 \\
\hline
0.5 & 142 & 5.2 & 0.79 & 1.31 \\
\hline
1.0 & 194 & 4.5 & 0.60 & 0.98 \\
\hline
1.5 & 296 & 3.7 & 0.40 & 0.66 \\
\hline
1.7 & 252 & 4.0 & 0.33 & 0.54 \\
\hline
1.9 & 205 & 4.4 & 0.27 & 0.44 \\
\hline
2.1 & 174 & 4.7 & 0.22 & 0.36 \\
\hline
2.24 & 174 & 5.3 & 0.19 & 0.31 \\
\hline
\end{tabular*}
\end{table}

\begin{table}[h]
\caption{The flight path, radius of last TPC hit, momentum resolutions, and TOF PID limits for pions and protons for the fixed-target events.} 
\label{FXT_TOF_table}
\begin{tabular*}{0.5\textwidth}{@{}l*{15}{@{\extracolsep{0pt
          plus12pt}}l}}
$\eta$ & flight path & $r_{\rm last\,hit}$ & $\delta p/p (\pi)$ & $\pi/K$ & $\delta p/p$(pro) & $(\pi+K)/p$ \\
 & (cm) & (cm) & (\%) & (GeV/c) & (\%) & (GeV/c) \\
\hline
0.0 & 220 & 200 & 1.6  & 1.7 & 2.8 & 2.6 \\
\hline
0.5 & 248 & 200 & 1.6  & 1.6 & 2.7  & 2.4 \\
\hline
1.0 & 339 & 200 & 1.5  & 1.3 & 2.4  & 2.1 \\
\hline
1.47 & 453 & 200 & 1.5  & 1.1 & 2.1  & 1.7 \\
\hline
1.52 & 539 & 188 & 1.5  & 1.1 & 2.1  & 1.7 \\
\hline
1.6 & 531 & 172 & 1.6  & 1.0 & 2.2  & 1.6 \\
\hline
1.7 & 523 & 155 & 1.7  & 0.9 & 2.4  & 1.4 \\
\hline
1.8 & 517 & 139 & 1.9  & 0.8 & 2.6  & 1.3 \\
\hline
1.9 & 512 & 125 & 2.0  & 0.74 & 2.8  & 1.2 \\
\hline
2.0 & 508 & 113 & 2.2  & 0.67 & 2.9  & 1.05 \\
\hline
2.1 & 505 & 102 & 2.4  & 0.61 & 3.1  & 0.95 \\
\hline
2.2 & 502 & 92 & 2.7  & 0.55 & 3.3  & 0.86 \\
\hline
\end{tabular*}

\end{table}

\subsection{Energy Range Accessible}
In fixed-target mode, the center-of-mass of the system is boosted in rapidity. 
Therefore, the mid-rapidity PID provided by the eTOF upgrade affects the range of 
energies that can be studied. Table~\ref{FXT_table} shows a listing of the proposed 
fixed-target energies and the corresponding rapidity offsets. The $y_{CM}$ resulting from the 
boosts are indicated in Figs. \ref{Acceptance_FXT_pro}, \ref{Acceptance_FXT_kap}, 
\ref{Acceptance_FXT_pip}, and \ref{Acceptance_FXT_ele} for the higher energies of the
fixed-target program. From these
figures, it is evident that the PID provided by the eTOF is needed for pion, kaon, and 
proton mid-rapidity studies at energies of 4.5, 5.2, 6.2, and 7.7 GeV.
It should be noted that even with the eTOF PID, studies of protons and kaons will be
limited to backward of mid-rapidity for the 7.7~GeV system. This system will also be 
studied in the collider program with a larger PID acceptance. Thus the 7.7 GeV system 
will provide important systematic consistency checks between fixed-target and collider 
mode, however, it is not necessary that all analyses 
be available for such checks. Without the eTOF,
the 5.2, 6.2, and 7.7~GeV fixed-target energies will not be run. These three energies require
RHIC to be tuned for energies that are not part of the BES-II program, and this can not be
justified without the mid-rapidity acceptance with PID. 
It should also be noted that if these energies are not run in fixed-target mode, 
a 170 MeV $\mu_B$ gap will be left in the energy scan just below the 215~MeV 
$\mu_B$ range covered by the collider portion of the scan (see Table~\ref{FXT_table}). 

\begin{table}[h]
\caption{The collider and fixed-target center-of-mass energies ($\sqrt{s_{NN}}$), projectile kinetic
energies (AGeV), center-of-mass rapidity offset ($y_{CM}$), and baryon chemical
potentials ($\mu_B$) for the proposed fixed-target program.}
\label{FXT_table}
\begin{tabular*}{0.5\textwidth}{@{}l*{15}{@{\extracolsep{0pt
          plus12pt}}l}}
Collider & Fixed Target & AGeV & $y_{CM}$ & $\mu_B$ (MeV)\\
\hline
62.4 & 7.7 & 30.3 & 2.10 & 420 \\
39   & 6.2 & 18.6 & 1.87 & 487 \\
27   & 5.2 & 12.6 & 1.68 & 541 \\
19.6 & 4.5 &  8.9 & 1.52 & 589 \\
14.5 & 3.9 &  6.3 & 1.37 & 633 \\
11.5 & 3.5 &  4.8 & 1.25 & 666 \\
9.1  & 3.2 &  3.6 & 1.13 & 699 \\
7.7  & 3.0 &  2.9 & 1.05 & 721 \\
\hline
\end{tabular*}
\end{table}

Although a detailed proposal for running the fixed-target program has not been 
finalized, the general concept has always been a key part of the BES-II proposal.
Based on the performance from the 2015 fixed-target test, it is clear that the number of 
events which can be recorded is limited by the DAQ rate, the expected store length, 
and the machine duty cycle. The estimate is that approximately 50 million events can be 
recorded per day of running, independent of energy. 

\subsection{Mapping out the Phase Space}
\begin{figure}
 \begin{center}
 \includegraphics[width=0.45\textwidth]{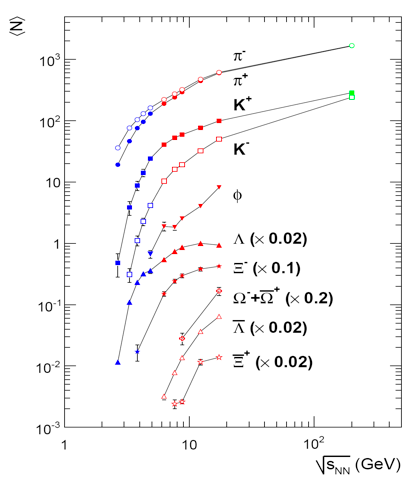}
 \caption{The yield per event of mesons, hyperons and anti-hyperons as a function of collision
energy, measured in central Au+Au or Pb+Pb collisions~\cite{Blume}.
Note that the cross sections of the hyperons and anti-hyperons are downscaled by various factors
in order to avoid overlapping plotting symbols.
} 
 \label{MultiStrangeYields}
 \end{center}
\end{figure}

Exploring the phase diagram of QCD matter requires that at each collision energy there is  
sufficient yield (both $y_{CM}=0$ and full acceptance) of each species to 
determine the chemical equilibrium $T$ and $\mu_B$ values. The coverage
maps shown in 
Figs.~\ref{Acceptance_FXT_pro}, \ref{Acceptance_FXT_kap}, and
\ref{Acceptance_FXT_pip} demonstrate that we have acceptance for $\pi$, $K$, and $p$ from
$y_{CM} = 0$ to $y_{\rm target}$ for all fixed-target energies except 7.7 GeV, where 
even with eTOF PID, the $K$ and $p$ acceptances do not reach $y_{CM} = 0$.
The efficiency for hyperon reconstruction is a convolution of the single
particle acceptances. This will make possible $y$-dependent measurements
of $K^0_S$, $\Lambda$, and $\Xi^-$. Currently, there is only a single $\Xi^-$ measurement
for collision energies below 7.7 GeV~\cite{Chung:2003zr}. The STAR fixed-target program will map out the turn on of $\Xi$
production with collision energy. Measurements of $\Omega$, $\bar{\Lambda}$, and 
$\bar{\Xi}^+$ have not been made at these energies previously 
(see Fig.~\ref{MultiStrangeYields}). Studying the onset of the production of these
species could be possible with the fixed-target program using the eTOF.

\subsection{The Onset of Deconfinement}
NA49 has argued that the onset of deconfinement is achieved 
at 7.7 GeV~\cite{Alt:2007aa}. 
This result is based on a set of inclusive observables: there is a kink in the rate of
increase of the pion production with collision energy, there is a step in the slope parameter of
the kaon spectra, and there is a peak (horn) in the $K^+/\pi^+$ ratio. We will study all of 
these inclusive observables. In addition, the fixed-target program will allow us to track  
other QGP signature observables, studied in BES-I, spanning a collision energy range 
from 3.0 to 19.6 GeV ($\mu_B$ from 720 to 205 MeV). 
The eTOF PID is essential for a number of deconfinement observables that will be studied in the fixed-target program:
\begin{itemize}
  \item{Number of constituent quark scaling of elliptic flow is a key QGP 
signature~\cite{Adams:2005zg}.
The results from BES-I show the $N_{CQ}$ scaling is exhibited independently for particles
and anti-particles~\cite{Adamczyk:2013gv}. At fixed-target energies, the $N_{CQ}$ 
scaling for particles is expected to break. The elliptic flow is a mid-rapidity observable,
and PID is necessary in order to observe the $N_{CQ}$ scaling. This study is only possible with 
eTOF as it is needed for mid-rapidity proton PID for all fixed-target energies except 3.0 GeV.} 
  \item{One form of balance function is a rapidity correlator. This correlator is sensitive to QGP 
formation~\cite{Pratt:2011bc}. The BES-I data show the balance function narrowing signal decreases with decreasing beam
energy. This signal is almost, but not quite, gone at 7.7 GeV~\cite{BES_Balance_Functions}. 
Lower energy measurements are needed to demonstrate that this signature disappears. A key
to the sensitivity of this measurement is the width of the total rapidity window. The eTOF
extends the rapidity range and will improve the significance of this study.}
  \item{Strangeness enhancement is seen as an important QGP signature. The energy range covered
by the fixed-target program sees the opening of several strange particle production channels
(see Fig.~\ref{MultiStrangeYields}). The strange particle production is maximum at mid-rapidity,
and cleanly identifying weakly decaying strange hadrons requires clean PID for the daughters. 
The eTOF significantly improves the efficiency and acceptance for strange hadrons and allows
measurements of their rapidity density distributions. } 
\end{itemize}

\subsection{Compressibility and the First-Order Phase Transition}
Assuming that there is a first-order phase transition, the concept of a single ``onset 
of deconfinement" is an oversimplification. Depending on the universality class 
of the phase transition, there may be a spinodal decomposition which would imply a 
mixed-phase region with a negative compressibility. Rather than a single ``onset", there  
may actually be several interesting onsets: the lowest energy which causes some fraction  
of the system to enter the mixed phase region, the energy at which the system spends the  
maximum amount of time in the instability regime, and the energy at which the system 
passes into the pure QGP phase. In order to understand the nature of the phase transition, 
we propose to study several observables which are expected to have sensitivity to  
the compressibility. The observables for which eTOF PID is essential include: 
\begin{itemize} 
\item{The directed flow of protons, which offers sensitivity to the early compressibility.
Most of these particles are transported participants recoiling off the interaction 
region~\cite{Adamczyk:2014ipa, Stoecker:2004qu, PhysRevC.89.054913, PhysRevC.90.014903,
phsd2, Nara:2016phs}. 
The eTOF is needed to study the evolution of the mid-rapidity``wiggle'' which is 
particularly sensitive to compressibility and is known to be absent~\cite{Liu:2000am}
at the lower end of the proposed fixed-target beam energy range.}
\item{The tilt angle of the pion source, measured through femtoscopy
~\cite{Lisa:2000ip,Lisa:2000xj,Lisa:2011na}. The eTOF is needed to identify 
mid-rapidity pions for fixed-target energies from 4.5 to 7.7 GeV.} 
\item{The volume of the pion source, measured through femtoscopy~\cite{Lisa:2005dd}. The
eTOF is needed for mid-rapidity PID for energies from 4.5 to 7.7 GeV.} 
\item{The width of the pion rapidity density distribution, which has been argued to be 
sensitive to the speed of sound in nuclear matter~\cite{Petersen:2006mp}. This
study relies on the ability to characterize the shape of the rapidity density
distribution. Measurements with sufficient rapidity acceptance are required in order to understand the
shape of the distribution. Mid-rapidity pion measurements are only possible using the
eTOF for PID for energies above 4.5 GeV.} 
\item{The elliptic flow of protons, which has been shown to change sign at a fixed-target 
beam energy of 6~AGeV ($\sqrt{s_{NN}} = 3.5$ GeV)~\cite{Pinkenburg:1999ya}. 
This sign change of $v_2$ is explained by the transit speed of the projectile 
nucleus through the target nucleus matching the expansion speed from compression 
(speed of sound). Mid-rapidity PID from eTOF is essential for further investigation 
of this behavior.} 
\item{The Coulomb potential of the pion source provides an independent means of 
assessing the source volume, being affected by the expansion velocity of the 
system~\cite{Baym96}. To study this physics, pion PID from eTOF is needed at mid-rapidity. } 
\item{The life-time of the emitting source, measured through low-mass 
dileptons~\cite{Rapp2016586}.}
\end{itemize}

\subsection{Criticality}
The observation of enhanced fluctuations would be the clearest evidence that the 
reaction trajectory of the cooling system had passed near the possible critical
end point on the QGP/hadronic gas phase boundary. Recent analyses of the 
higher moments of the net-proton distributions have suggested the possibility of 
enhanced fluctuations at 7.7 GeV~\cite{Adamczyk:2013dal}. 
These results require higher statistics to improve the significance. 
The significance of the higher moments signal has also been shown to depend strongly
on acceptance. It was shown in Fig.~\ref{Kurtosis_etof} that the significance
of the $\kappa \sigma^2$ signal scales as $N_p^3$.

An important test to determine if the enhanced 
fluctuations are related to critical behavior would be to see the fluctuation signals 
return to their base-line levels at energies below 7.7 GeV. The energies of the 
fixed-target program will provide these important control studies, allowing critical 
behavior searches to be extended to higher $\mu_B$. Clean PID is essential for this
analysis and its sensitivity has been shown to depend strongly on the efficiency and
acceptance. At 4.5 GeV, test run data indicate that we accept 20\% of all protons
using the current configuration of the TPC. This would drop to 5\% for the 7.7~GeV
fixed-target energy. The key mid-rapidity coverage of the eTOF raises these acceptance
numbers to 50\% and 20\% respectively (increasing the significance of the results
by a factor of 15 to 65). Although there are some 
fluctuation analyses performed by the NA49~\cite{Afanasev:2000fu} 
collaboration, the more sensitive higher moments studies have been done only by 
STAR~\cite{Adamczyk:2013dal,Adamczyk:2014fia}, PHENIX~\cite{Adare:2015aqk}, 
and are being studied by HADES at 2.42 GeV. 
There were no critical fluctuation studies performed at the AGS so the fixed
target program will provide the first data in this energy regime.

\subsection{Chirality}
Dilepton experiments have been an important part of the physics program 
at almost all heavy-ion facilities, with the notable exception of the AGS. 
At the lowest energies, HADES measured e+e- productions in Au+Au collisions at
$\sqrt{s_{NN}}$ = 2.42 GeV. In the SPS heavy-ion program, 
dilepton data were taken by experiments Helios-3, NA38/50, CERES, and NA60. 
And at RHIC, both PHENIX and 
STAR have dilepton capabilities.  
The eTOF detector will provide electron ID at mid-rapidity for all energies of the 
fixed-target program. This provides the first opportunity to study the evolution of  
the low-mass dilepton excess in this collision energy region in which the low-mass 
yield might be also sensitive to the emitting source 
temperature in addition to being sensitive 
to the total baryon density.
These dependencies will help us to understand the mechanism of in-medium $\rho$ broadening 
which is the fundamental probe of chiral symmetry restoration in hot, dense QCD matter.

\subsection{Hypernuclei}
The first species of hypernuclei, $^3_\Lambda$H and $^4_\Lambda$H, were  
discovered in the 1950s~\cite{Pniewski}. Several isotopes of hyper-helium and 
hyper-lithium have been found in kaon beam $s$-transfer reactions. In heavy-ion 
collisions, light nuclei 
are formed through coalescence of nucleons. As the energy is raised, nucleons can 
coalesce with hyperons to form light hypernuclei. At even higher energies,   
anti-nucleons can coalesce to form light anti-nuclei. This coalescence mechanism has 
allowed STAR to make the discoveries of anti-hyper-tritium 
($\overline{^3_\Lambda \mathrm{H}}$)~\cite{Abelev:2010rv} and 
anti-alpha ($\overline{^4 \mathrm{He}}$)~\cite{Agakishiev:2011ib}. 

\begin{figure}
 \begin{center}
 \includegraphics[width=0.45\textwidth]{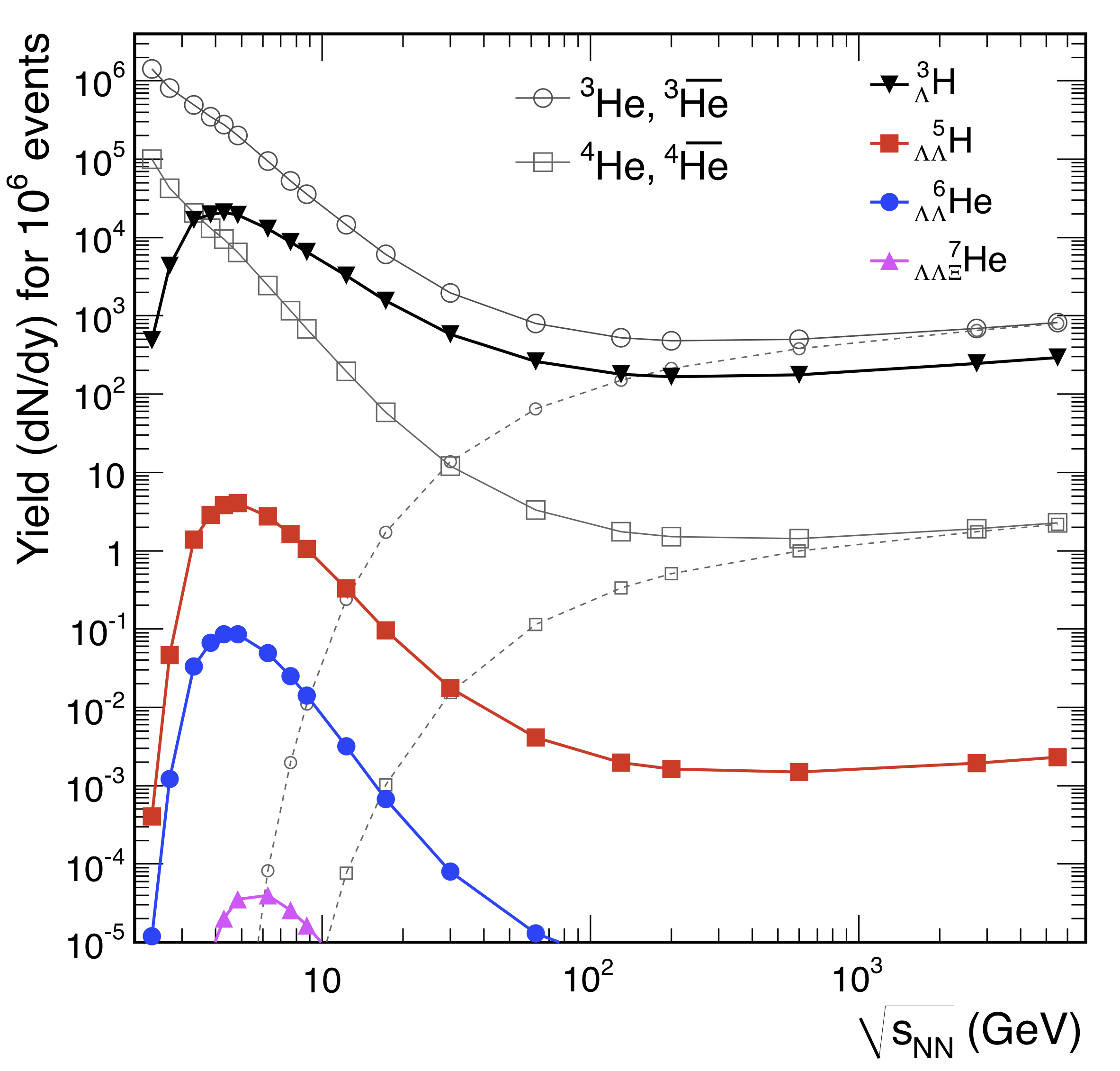}
 \caption{The energy dependence of hypernuclei yields at mid-rapidity in Au+Au collisions
          calculated using the statistical model of Ref. \cite{Andronic:2010qu}. } 
 \label{Hypernuclei}
 \end{center}
\end{figure}

\begin{figure}
 \begin{center}
 \includegraphics[width=0.45\textwidth]{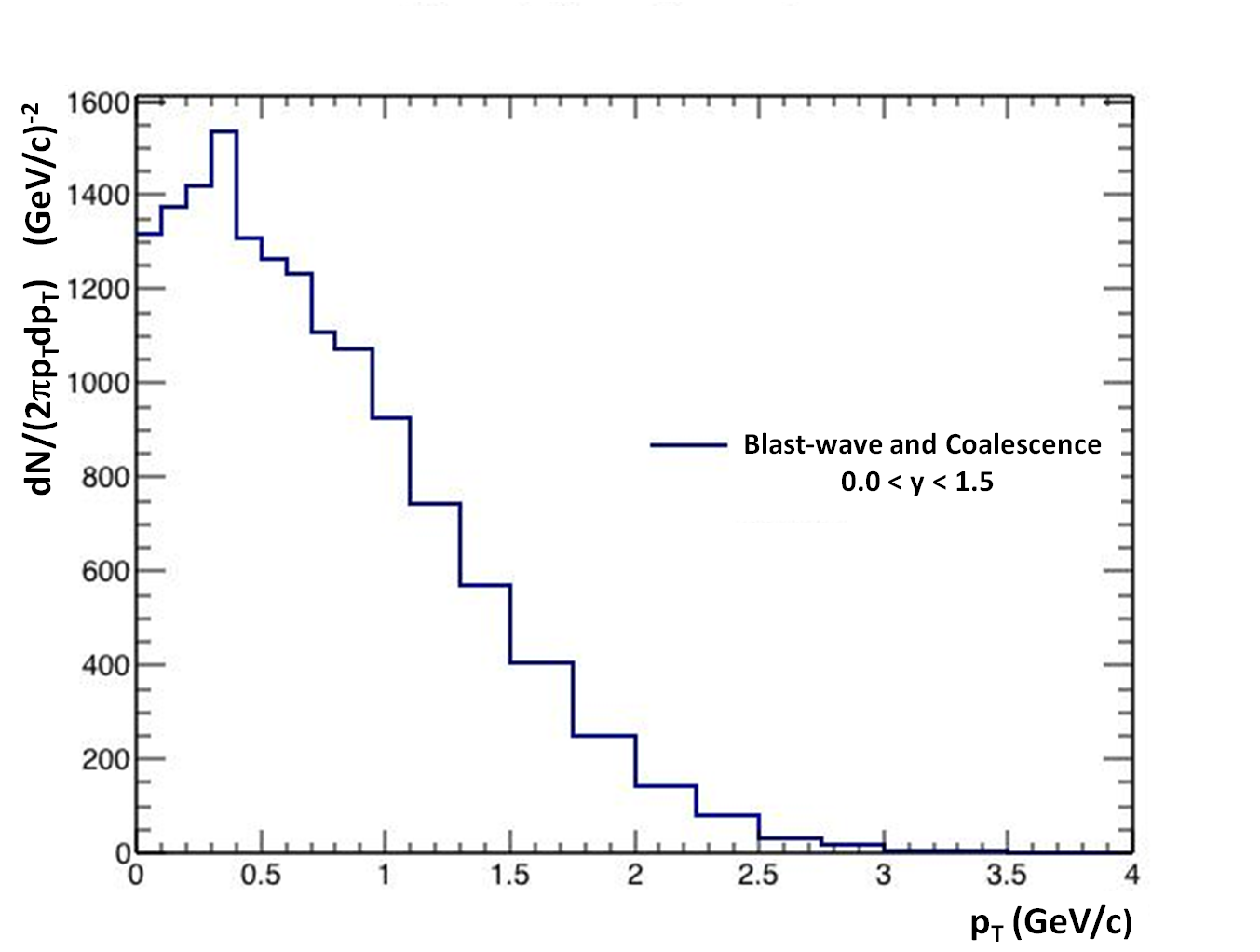}
 \caption{The simulated $p_T$ distribution of hypertritons from one day of running
for fixed-target Au+Au collisions at 4.5~GeV.} 
 \label{hyperT_pt}
 \end{center}
\end{figure}

 The energy regime covered by the fixed-target program (3.0 to 7.7 GeV) should be optimal 
for the formation of matter (as opposed to anti-matter) hypernuclei. 
At energies below 3.0 GeV, few hyperons are produced, whereas at energies above 8 GeV,  
the increased production of anti-baryons stifles formation of composites of matter particles
(see Fig.~\ref{Hypernuclei}). Meaningful samples of $^3_\Lambda$H and $^4_\Lambda$H will 
be observed at all the fixed-target energies. Figure \ref{hyperT_pt} shows the 
expected $p_T$ distribution of hypertritons from a single day of running at 4.5 GeV
assuming a coalescence model for production and a blast wave model to describe the 
$p_T$ distribution. 
These simulations were made assuming eTOF coverage and PID, which increases the 
efficiency of $^3_\Lambda$H reconstruction by a factor of eight.
The statistics are expected to be comparable to STAR data samples from 200 GeV collider data; 
this will allow a precise measurement of the light hypernuclei lifetimes and a 
mapping of the $^3_\Lambda$H$/(^3$He$ \times (\Lambda/p))$,  
and $^4_\Lambda$H$/(^4$He$ \times (\Lambda/p))$ 
ratios as a function of $\sqrt{s_{NN}}$. Searches for multi-strange hypernuclei  
($^5_{\Lambda\Lambda}$H and $^6_{\Lambda\Lambda}$He) make appealing physics goals.  
However, further simulations are required to determine if these measurements will be feasible with 
the expected integrated luminosity.

\section{Summary}
The eTOF upgrade to the STAR detector complements the iTPC upgrade and enables forward 
PID, critical for precision studies of the rapidity dependence of key bulk property observables.
Because this energy regime is characterized by the incomplete transparency of the participant 
nucleons (partial stopping), inspecting the particle distributions over a wide rapidity interval 
will help to disentangle dynamic from thermodynamic effects. This 
additional analysis handle will help constrain the models and help clarify the
phase diagram of QCD matter.  We emphasize the following signals: 
\begin{itemize} 
\item{Dileptons. Extending the rapidity for PID will provide an independent observable to study 
the baryon density dependence of low-mass dielectron emission, extend electron ID to $\eta = 1.5$, and   
quantify the baryon density effect on $\rho$ broadening.}
\item{Directed flow. Extending PID to $y$ = 1.2 will allow the study of $v_1$ over a new rapidity region. 
This can help confirm the possible softening of the equation of state.}
\item{Elliptic flow. Rapidity dependent measurements of $v_2$ will be enabled.  The $\phi$ flow can be 
studied.}
\item{Fluctuations. Enhanced fluctuation signals will provide a cleaner and more significant 
indication of possible critical behavior.}
\end{itemize}
For the internal fixed-target program, the eTOF will provide PID for mid-rapidity and justify 
running at collision energies from 4.5 to 7.7 GeV. 
This will make it possible to have a 
comprehensive scan from $\sqrt{s_{NN}}$ = 3.0 to 19.6~GeV, $\mu_B$ = 720 to 205 MeV. 
Without the mid-rapidity PID provided by eTOF, there would be a large gap in
the data from $\mu_B$ = 590 to 420 MeV. 
The energy range from 3.0 to 19.6 GeV spans from regions well understood to be compressed baryonic matter up
to regions for which partonic behavior is well established.
The eTOF upgrade to the STAR detector will bring compelling new physics capabilities 
to the RHIC BES-II program.

\pagebreak

\bibliography{eTOF_Physics}

\end{document}